\pdfoutput=1
\documentclass[11pt]{article}
\usepackage[a4paper,margin=1in]{geometry}
\usepackage{amsmath,amssymb}
\usepackage[T1]{fontenc}
\usepackage[utf8]{inputenc}
\usepackage{lmodern}
\usepackage{microtype}
\usepackage{booktabs}
\usepackage{graphicx}
\usepackage[hidelinks]{hyperref}

\numberwithin{equation}{section}

\IfFileExists{generated/fit_tables.tex}{\newcommand{\FivLimit}{0.015}
\newcommand{\FivSlope}{205}
\newcommand{\FivDeltaOne}{2.9}
\newcommand{\FivDeltaTwo}{7.4}
\newcommand{\FivDeltaThree}{13.7}
\newcommand{\FivDeltaFour}{26.1}
\newcommand{\FivFitResid}{0.02}
\newcommand{\ParTableRows}{$\epsilon_{\mathrm{nd}}$ & the non-diffractive mass law $(M^2)^{-1-\epsilon_{\mathrm{nd}}}$ & $0.2210 \pm 0.0075$ \\
$b$ [GeV$^{-2}$] & the vertex slope at the proton mass & $5.64 \pm 0.09$ \\
$f$ & the intact proton's vertex term & $1.26 \pm 0.04$ \\
$s_{\mathrm{tail}}$ & the leading-proton tail's share of single diffraction & $0.218 \pm 0.010$ \\
$s_{\pi}$ & the one-pion exchange's share of single diffraction & $0.272 \pm 0.026$ \\
$\gamma_s$ & the weight of a created $s\bar s$ pair & $0.267 \pm 0.011$ \\
$r_V$ & the vector against pseudoscalar weight & $0.99 \pm 0.05$ \\
$r_{B\bar B}$ & the weight of a baryon pair at a continuing step & $3.62 \pm 0.16$ \\
$a_{YK}$ & the associated hyperon--kaon weight at a non-strange baryonic vertex & $1.00 \pm 0.07$ \\
$F_B$ & the leading-baryon split share & $0.166 \pm 0.006$ \\
$F_M$ & the leading-meson split share & $0.0442 \pm 0.0033$ \\
$\kappa_B$ & the baryon split's direction law & $1.66 \pm 0.19$ \\
$b_M$ [GeV$^{-2}$] & the meson split's transverse slope & $2.76 \pm 0.43$ \\
$f_{\mathrm{iv}}$ & the isovector exchange fraction & $< 0.015$ (95\%) \\}
\newcommand{\NFitParamsOther}{thirteen}
\newcommand{\NFitParams}{fourteen}
\newcommand{\NFitParamsCap}{Fourteen}

\newcommand{\FitChiSq}{2263}
\newcommand{\FitNdf}{1248}
\newcommand{\FitSsq}{1.81}
\newcommand{\FitS}{1.35}
\newcommand{\ClassRows}{non-diffractive & 0.577 & 18.34 \\
single diffraction, projectile & 0.188 & 5.97 \\
single diffraction, target & 0.188 & 5.97 \\
double diffraction & 0.047 & 1.50 \\}
\newcommand{\SDTotalMb}{11.94}
\newcommand{\SDWindowMb}{4.90}
\newcommand{\OPEMbPerSide}{1.63}
\newcommand{\TailMbPerSide}{1.30}
\newcommand{\OPEUnabsorbedMb}{3.06}
\newcommand{\OPEAbsorbedMb}{1.49}
\newcommand{\OPEUnabsorbedShare}{0.51}
\newcommand{\OPEAbsorbedShare}{0.25}
\newcommand{\OPEOverUnabsorbed}{0.53}
\newcommand{\OPEOverAbsorbed}{1.09}
\newcommand{\ParEpsNd}{0.221 \pm 0.007}
\newcommand{\ValEpsNd}{0.221}
\newcommand{\ParSPi}{0.272 \pm 0.026}

\newcommand{\ParGammaS}{0.267 \pm 0.011}

\newcommand{\ParRV}{0.99 \pm 0.05}

\newcommand{\ParRBB}{3.62 \pm 0.16}

\newcommand{\ParAYK}{1.00 \pm 0.07}

\newcommand{\GammaSRelation}{below one}
\newcommand{\RVRelation}{within one error of one}
\newcommand{\RBBRelation}{above one}
\newcommand{\AYKRelation}{within one error of one}
\newcommand{\KplusIdPull}{2.0}
\newcommand{\KminusIdPull}{3.7}
\newcommand{\KminusIdLift}{22}
\newcommand{\NormPullProton}{-2.24}
\newcommand{\NormPullAntiproton}{0.31}
\newcommand{\NeutronScalePct}{-10}
\newcommand{\NeutronScalePull}{1.0}
\newcommand{\NormWithinOne}{five}
\newcommand{\AntiprotonYieldRatio}{0.96}

\newcommand{\SegmentRows}{$\pi^-$ & 281 & 447 & 1.59 & 0.962 \\
$\pi^+$ & 260 & 347 & 1.34 & 0.993 \\
$K^+$ & 153 & 185 & 1.21 & 0.980 \\
$K^-$ & 154 & 268 & 1.74 & 0.980 \\
$p$ & 284 & 808 & 2.85 & 1.031 \\
$\bar p$ & 122 & 180 & 1.47 & 0.955 \\
$n$ & 8 & 3 & 0.33 & -- \\}
\newcommand{\ChiPerPointKplus}{1.21}
\newcommand{\ChiPerPointKminus}{1.74}
\newcommand{\ChiPerPointAntiproton}{1.47}
\newcommand{\ChiPerPointPionProtonRange}{1.3 to 2.8}
\newcommand{\NuisanceRows}{normalization $\pi^-$ (1.5\%) & +1.4\% & $+0.97$ \\
normalization $\pi^+$ (1.5\%) & -0.6\% & $-0.40$ \\
normalization $K^+$ (1.5\%) & -0.8\% & $-0.54$ \\
normalization $K^-$ (1.5\%) & +0.3\% & $+0.20$ \\
normalization $p$ (1.5\%) & -3.4\% & $-2.24$ \\
normalization $\bar p$ (1.5\%) & +0.5\% & $+0.31$ \\
$K^+$ identification, $x_F \ge 0.25$ (4--12\%) & +1.98 amplitudes & $+1.98$ \\
$K^-$ identification, $x_F \ge 0.25$ (0--6\%) & +3.73 amplitudes & $+3.73$ \\
neutron scale (10\%) & -10.3\% & $-1.03$ \\}
\newcommand{\NuisancePenalty}{25.5}
\newcommand{\NeutronRows}{0.10 & $0.481 \pm 0.100$ & 0.462 & 0.96 \\
0.20 & $0.407 \pm 0.060$ & 0.438 & 1.08 \\
0.30 & $0.378 \pm 0.050$ & 0.423 & 1.12 \\
0.40 & $0.325 \pm 0.037$ & 0.373 & 1.15 \\
0.50 & $0.325 \pm 0.040$ & 0.335 & 1.03 \\
0.60 & $0.293 \pm 0.030$ & 0.331 & 1.13 \\
0.75 & $0.286 \pm 0.030$ & 0.341 & 1.19 \\
0.90 & $0.215 \pm 0.060$ & 0.235 & 1.09 \\}
\newcommand{\VariantRows}{discrepancy floor 3\% & +1601 \\
discrepancy floor 7\% & -746 \\
discrepancy floor 10\% & -1309 \\
common normalization shift $+2\%$ & +3 \\
common normalization shift $-2\%$ & +9 \\
$\sigma_{DD}$ 0.5 mb & +29 \\
$\sigma_{DD}$ 2.5 mb & -11 \\
$\sigma_{SD}(M^2/s<0.05)$ 4.1 mb & +75 \\
$\sigma_{SD}(M^2/s<0.05)$ 5.7 mb & -15 \\}
\newcommand{\GeneratorRows}{this model & 447 & 347 & 185 & 268 & 808 & 180 & 2260 \\
PYTHIA 8 (SoftQCD) & 3141 & 5091 & 303 & 299 & 20897 & 220 & 32430 \\
EPOS-LHC & 2051 & 2771 & 1617 & 399 & 2294 & 175 & 9478 \\
UrQMD 4 & 3298 & 4131 & 468 & 367 & 13983 & 2217 & 25050 \\}
\newcommand{\GeneratorSumRows}{this model & 0.95 & 0.99 & 0.99 & 0.99 & 1.01 & 0.97 \\
PYTHIA 8 (SoftQCD) & 1.07 & 1.01 & 1.26 & 1.49 & 0.94 & 2.46 \\
EPOS-LHC & 1.05 & 1.05 & 1.07 & 1.01 & 1.07 & 1.04 \\
UrQMD 4 & 0.86 & 0.84 & 0.94 & 1.06 & 0.93 & 1.20 \\}
\newcommand{\GeneratorPullRows}{this model & $+0.97$ & $-0.40$ & $-0.54$ & $+0.20$ & $-2.24$ & $+0.31$ & $+1.98$ & $+3.73$ & 24 \\
PYTHIA 8 (SoftQCD) & $-8.05$ & $-4.55$ & $-13.5$ & $-20.4$ & $-20.2$ & $-36.9$ & $+4.46$ & $+2.50$ & 2478 \\
EPOS-LHC & $+0.49$ & $+1.10$ & $-7.98$ & $-1.83$ & $-7.69$ & $-2.88$ & $+4.94$ & $-3.16$ & 170 \\
UrQMD 4 & $+1.76$ & $+6.88$ & $+3.37$ & $-4.18$ & $-14.3$ & $-17.4$ & $-0.29$ & $-1.25$ & 586 \\}
\newcommand{\GeneratorSpeciesSums}{29{,}952, 9{,}307, 24{,}464}
\newcommand{\CascadeSpeciesSum}{2{,}235}
\newcommand{\GeneratorEvents}{3.2}
\newcommand{\RhoZero}{0.349}
\newcommand{\OmegaYield}{0.496}
\newcommand{\PhiYield}{0.0160}
\newcommand{\KstarZero}{0.0994}
\newcommand{\KstarZeroBar}{0.0557}

\newcommand{\RhoOverOmega}{0.70}
\newcommand{\KstarRatio}{1.78}
\newcommand{\XiMinusModel}{0.00202}
\newcommand{\XiMinusRatio}{0.61}
\newcommand{\XiBarModel}{0.001154}
\newcommand{\XiBarRatio}{1.46}
\newcommand{\XiLineData}{0.0025}
\newcommand{\XiLineRatio}{0.35}
\newcommand{\OmegaMinusModel}{6.0}
\newcommand{\OmegaBarModel}{3.8}
\newcommand{\OmegaBarRatio}{0.63}

\newcommand{\NchLowRange}{0.91 to 1.03}
\newcommand{\NchAtTwoSix}{3.6}
\newcommand{\NchAtTwoSixRef}{3.7}

\newcommand{\NchAtFive}{5.9}
\newcommand{\NchAtFiveRef}{4.8}

\newcommand{\NchAtSevenEight}{9.1}
\newcommand{\NchAtSevenEightRef}{5.5}
\newcommand{\NchAtSevenEightRatio}{1.65}

\newcommand{\NchSlope}{1.13}
\newcommand{\SdLawInside}{1.07 to 1.17}

\newcommand{\SdLawBeyondText}{1.04 times its continuation at $M_X$ of 6--8~GeV, 1.18 at 8--10~GeV, 1.47 at 10--13~GeV, and 1.74 at 13--17.3~GeV}
\newcommand{\SdTopEvent}{14.0}
\newcommand{\SdTopRatio}{1.7}
\newcommand{\NchEventModel}{6.93}
\newcommand{\NchEventData}{6.94}
\newcommand{\SeagullRows}{$\pi^-$ & 16 & 0.259 & 0.261 & 0.018 & 0.021 & 0.020 & 0.094 & +0.005 & +0.012 & -0.007 & +0.061 \\
$\pi^+$ & 12 & 0.260 & 0.260 & 0.003 & 0.029 & 0.029 & 0.041 & -0.001 & +0.028 & -0.024 & +0.026 \\
$K^+$ & 9 & 0.418 & 0.407 & 0.006 & 0.005 & 0.050 & 0.029 & -0.002 & -0.001 & -0.049 & -0.002 \\
$K^-$ & 10 & 0.402 & 0.394 & 0.036 & 0.034 & 0.026 & 0.054 & +0.023 & +0.013 & -0.014 & +0.028 \\
$p$ & 14 & 0.515 & 0.516 & 0.014 & 0.022 & 0.023 & 0.056 & -0.011 & -0.017 & -0.019 & +0.041 \\
$\bar p$ & 9 & 0.490 & 0.492 & 0.018 & 0.024 & 0.016 & 0.087 & +0.015 & -0.023 & +0.006 & +0.085 \\}
\newcommand{\SeagullKminusFwd}{0.04 to 0.09}
\newcommand{\SeagullInterceptPion}{0.26}
\newcommand{\SeagullInterceptProton}{0.52}
\newcommand{\SeagullInterceptMaxDev}{0.010}
\newcommand{\SeagullRmsPiPlus}{0.003}
\newcommand{\SeagullRmsKPlus}{0.006}
\newcommand{\SeagullRmsPiMinus}{0.018}
\newcommand{\SeagullRmsAntiproton}{0.018}
\newcommand{\SeagullPimFwdXf}{0.85}
\newcommand{\SeagullPimFwd}{0.07}
\newcommand{\SeagullProtonBelow}{0.02}
\newcommand{\MesonSplitPct}{4.4}
\newcommand{\BaryonSplitPct}{16.6}
\newcommand{\BaryonSplitShare}{0.166}
\newcommand{\ValFM}{0.0442}
\newcommand{\MesonSplitRemainder}{0.956}
\newcommand{\BaryonBranchPct}{15.9}
\newcommand{\OrdinaryBranchPct}{79.7}
\newcommand{\NdDualityMass}{2.2}
\newcommand{\DiffDualityMass}{1.7}
\newcommand{\RedrawMass}{1.2}
\newcommand{\ContinuationFloor}{2.0}

\newcommand{\SplitKappa}{1.7}
\newcommand{\MesonSplitSlope}{2.8}
\newcommand{\NeutralOPEWindowPct}{0.9}
\newcommand{\EpsFree}{0.027 \pm 0.031}
\newcommand{\EpsFreePull}{1.8}
\newcommand{\AlignKappa}{1.5}
\newcommand{\AlignGain}{582}

\newcommand{\RedrawOffCost}{169}
\newcommand{\VertexExpLowCost}{$+575 \pm 47$}
\newcommand{\VertexExpHighCost}{$+624 \pm 57$}
\newcommand{\HeavyClusterCost}{$+22{,}700$}
\newcommand{\HeavyStringCost}{$+42{,}500$}
\newcommand{\HeavyTwoLineCost}{$+5{,}200$}
\newcommand{\HeavyMassScaleCost}{$+20{,}600$}
\newcommand{\NoHyperonSplitCost}{236 to 287}
\newcommand{\ReggeRemainderCost}{659 to 745}

\newcommand{\FixedChoiceRows}{diffractive intercept $\epsilon$ & 0.0808 & Donnachie--Landshoff~\cite{donnachie1992totalxsec} & fitted free: $0.027 \pm 0.031$ \\
Pomeron slope $\alpha'_{\mathbb{P}}$ & 0.25~GeV$^{-2}$ & Donnachie--Landshoff~\cite{donnachie1992totalxsec} & -- \\
coherence edge $M^2/s$ & 0.1 & Goulianos~\cite{goulianos1983diffractive} & -- \\
$\sigma_{SD}(M^2/s < 0.05)$ & 4.9~mb & measured~\cite{schamberger1975} & variants $\pm 0.8$~mb \\
$\sigma_{DD}$ & 1.5~mb & estimate~\cite{goulianos2004wherestand} & variants 0.5, 2.5~mb \\
continuum density $T_H$ & 165~MeV & Hagedorn~\cite{hagedorn1965thermo} & -- \\
vertex-slope exponent & 1.25 & the data & 1.0: $+575 \pm 47$; 1.5: $+624 \pm 57$ \\
continuing-step alignment $\kappa_L$ & 1.5 & the data & 1.0: $+105 \pm 23$; 2.0: $-22 \pm 20$; 3.0: $+65 \pm 30$; isotropic: $+582 \pm 39$; every step: $+120 \pm 31$ \\
non-diffractive duality boundary & 2.2~GeV & the data & 1.8: $+220 \pm 26$; 2.0: $+112 \pm 22$; 2.4: $+126 \pm 35$; 2.6: $+1{,}122 \pm 77$ \\
diffractive duality boundary & 1.7~GeV & the data & 1.6: $+2 \pm 8$; 1.8: $+5 \pm 7$ \\
redraw support & 1.2~GeV & the data & 1.1: $+169 \pm 12$; 1.15: $+46 \pm 7$; 1.3: $+54 \pm 7$; 1.4: $+98 \pm 9$; 1.5: $+290 \pm 16$ \\
continuation floor & 2.0~GeV & the data & 1.8: $+2{,}434 \pm 103$; 2.2: $+3{,}114 \pm 137$ \\
step minimum & 0.25~GeV/$c$ & the data & 0.15: $+649 \pm 54$; 0.35: $+388 \pm 55$ \\}
\newcommand{\RegPHardFwd}{1.00}
\newcommand{\RegPHardFwdN}{30}
\newcommand{\RegPHardFwdChi}{79}
\newcommand{\RegPSoftMid}{1.00}

\newcommand{\RegPSoftVfwd}{0.82}
\newcommand{\RegPSoftVfwdN}{17}
\newcommand{\RegPSoftVfwdChi}{118}
\newcommand{\RegPMid}{0.87}
\newcommand{\RegPMidN}{16}
\newcommand{\RegPMidChi}{120}
\newcommand{\RegPHardCentA}{0.82}

\newcommand{\RegPHardCentB}{0.94}

\newcommand{\RegPHardCentN}{9}
\newcommand{\RegPHardCentChi}{49}
\newcommand{\RegPimSoftMid}{1.06}
\newcommand{\RegPimSoftMidN}{15}
\newcommand{\RegPimSoftMidChi}{34}
\newcommand{\RegPimSoftCent}{0.95}
\newcommand{\RegPimSoftCentN}{25}
\newcommand{\RegPimSoftCentChi}{30}
\newcommand{\RegPipMid}{1.13}
\newcommand{\RegPipMidN}{9}
\newcommand{\RegPipMidChi}{77}
\newcommand{\RegPimVfwdA}{0.76}

\newcommand{\RegPimVfwdB}{0.91}

\newcommand{\RegPimVfwdN}{11}
\newcommand{\RegPimVfwdChi}{57}
\newcommand{\RegPipVfwd}{0.79}

\newcommand{\RegKpHardCentA}{0.76}

\newcommand{\RegKpHardCentB}{0.85}

\newcommand{\RegKpHardCentN}{10}
\newcommand{\RegKpHardCentChi}{50}
\newcommand{\RegKpBareA}{0.87}

\newcommand{\RegKpBareB}{0.78}

\newcommand{\RegKmLowA}{0.79}

\newcommand{\RegKmLowB}{0.76}

\newcommand{\RegKmHigh}{1.41}

\newcommand{\RegKmRowHigh}{1.23}

\newcommand{\RegKmFwdN}{25}
\newcommand{\RegKmFwdChi}{158}
\newcommand{\RegKpCent}{0.99}
\newcommand{\RegKpCentN}{101}

\newcommand{\RegKmCent}{1.00}

\newcommand{\RegKmRowLowRange}{0.45 to 0.62}

\newcommand{\PimVfwdRange}{0.76 to 0.91}
\newcommand{\KmFwdLowRange}{0.4 to 0.8}
\newcommand{\KmFwdHigh}{1.4}
\newcommand{\KmSegChi}{268}
\newcommand{\PVfwdCellRange}{0.77 to 0.95}
\newcommand{\KpLiftRange}{0.98 to 1.08}
\newcommand{\KpPipRatio}{0.88}
\newcommand{\NeutronRatioLow}{0.96}
\newcommand{\NeutronRatioMid}{1.15}
\newcommand{\NeutronRatioMax}{1.19}
\newcommand{\NeutronRatioMaxXf}{0.75}
\newcommand{\NeutronChi}{3}
\newcommand{\MCSpreadMedianPct}{13}
\newcommand{\MCSpreadRangePct}{12\% to 15}
\newcommand{\MCOverSigma}{0.21}
\newcommand{\MCDeltaChi}{36}
\newcommand{\MCSChangePct}{1}
\newcommand{\BCDataMedian}{0.995}
\newcommand{\BCDataDepartPct}{1.2}
\newcommand{\BCDataOverFive}{197}
\newcommand{\BCDataMaxPct}{20}
\newcommand{\BCModelChangePct}{0.2}
\newcommand{\BCModelOverFive}{37}
\newcommand{\BCModelBaryons}{32}
\newcommand{\BCModelChi}{174}
\newcommand{\VarFloorMaxMove}{1.9}
\newcommand{\VarNormMaxMove}{0.7}
\newcommand{\VarXsecMaxMove}{3.4}
\newcommand{\VarXsecMovedPars}{$f$ and $s_{\rm tail}$}
\newcommand{\NormScaleRange}{0.3\% to 3.4}
\newcommand{\KpLiftPctRange}{8\% to 24}
\newcommand{\GenRatioRange}{4.2 to 14.4}
\newcommand{\GenRatioRangeRound}{4 to 14}
\newcommand{\GenRatioEpos}{4.2}
\newcommand{\PhiRatioRange}{1.27 to 1.33}
\newcommand{\PhiSigmaNAfortynine}{2.7}
\newcommand{\LambdaRatios}{1.13, 1.27, 1.26, and 1.15}
\newcommand{\NoHyperonLambdaFirst}{1.23}
\newcommand{\NoHyperonLambdaLast}{1.01}

\newcommand{\LeverBaseLambdaRange}{1.16 to 1.29}
\newcommand{\ReggeLambdaRange}{1.17 to 1.35}
\newcommand{\LambdaTotal}{1.20}
\newcommand{\KstarBinRange}{1.10 to 1.61}
\newcommand{\KstarFwdBinRange}{1.37 to 1.61}
\newcommand{\KstarBarBinRange}{0.95 to 1.17}
\newcommand{\FwdExcessRange}{1.1 to 1.6}

\newcommand{\GenRatioUrqmd}{11.1}
\newcommand{\GenRatioPythia}{14.4}

\newcommand{\YieldWithinPct}{5}

\newcommand{\ESSMin}{0.791}
\newcommand{\ESSMax}{0.792}
\newcommand{\RegenMaxMove}{1.8}
\newcommand{\RegenSamples}{five}
\newcommand{\RegenScatter}{26}
\newcommand{\RegenReplicaMaxDiff}{1.7}
\newcommand{\RegenPairSpread}{40}
\newcommand{\SpecFile}{dev/mixing2006/config/runs/r146.json}
\newcommand{\RecordFile}{dev/00runs/paper\_record\_r146/RECORD.json}
}{}
\providecommand{\RhoZero}{--}\providecommand{\OmegaYield}{--}\providecommand{\PhiYield}{--}
\providecommand{\KstarZero}{--}\providecommand{\KstarZeroBar}{--}\providecommand{\RhoOverOmega}{--}\providecommand{\KstarRatio}{--}
\providecommand{\NchLowRange}{--}\providecommand{\NchAtTwoSix}{--}\providecommand{\NchAtTwoSixRef}{--}\providecommand{\NchAtFive}{--}\providecommand{\NchAtFiveRef}{--}\providecommand{\NchAtSevenEight}{--}\providecommand{\NchAtSevenEightRef}{--}\providecommand{\NchAtSevenEightRatio}{--}\providecommand{\NchSlope}{--}\providecommand{\SdLawInside}{--}\providecommand{\SdLawBeyondText}{--}\providecommand{\NchEventModel}{--}\providecommand{\NchEventData}{--}\providecommand{\SeagullKminusFwd}{--}\providecommand{\SdTopEvent}{--}\providecommand{\SdTopRatio}{--}\providecommand{\SeagullRows}{}\providecommand{\GeneratorRows}{}\providecommand{\GeneratorSumRows}{}\providecommand{\GeneratorEvents}{3.2}

\title{HEART: a resonance-cascade model of soft hadron production in p+p collisions at 158~GeV/$c$}
\author{Ondrej Chvala\thanks{This work is an independent phenomenological reanalysis of published NA49 data and is not an official publication of the NA49 Collaboration.}\\
\small Nuclear Engineering Teaching Laboratory, The University of Texas at Austin\\
\small Department of Nuclear Engineering, The University of Tennessee, Knoxville\\
\small \texttt{ochvala@utexas.edu}}
\date{November 2026}

\begin{document}
\maketitle

\begin{abstract}
Fischer and co-workers interpret the NA49 identified-hadron spectra in p+p collisions at 158~GeV/$c$ as the sum of a target and a projectile component, each the decay of an excited hadronic system. This paper supplies a complete event-level realization of that picture, the HEART model (Hadronic Excitations And Resonance Transitions), and tests it on the published NA49 cross-section data. Events are divided into four exchange classes, namely single diffraction on either side, double diffraction, and non-diffractive production. Each class produces color-singlet excited systems. Each system is a catalogued resonance below a duality boundary of \DiffDualityMass{} to \NdDualityMass~GeV, or a member of a resonant continuum above it. It de-excites by successive two-body transitions with exact conservation of four-momentum, baryon number, charge, and strangeness. The one-pion exchange of single diffraction produces the leading neutrons. \NFitParamsCap{} parameters are fitted to the 1,254 measured points of the $\pi^\pm$, $K^\pm$, $p$, and $\bar p$ double-differential cross sections on $0 \le x_F \le 0.95$ and $0.05 \le p_T \le 1.5$~GeV/$c$ alongside eight neutron $x_F$ points. Their errors are conditional on the model's discrete choices, which were fixed by fitting the alternatives to the same simulated events and comparing their $\chi^2$. Under an error model that profiles correlated systematic uncertainties, the fit yields $\chi^2 = \FitChiSq$ on \FitNdf{} degrees of freedom. The total yields lie within \YieldWithinPct\% of the measured ones. The model has no central source, and its isovector exchange fraction is consistent with zero. On the same points, the fixed-tune generators EPOS-LHC, UrQMD, and PYTHIA~8 produce $\chi^2$ values \GenRatioRange{} times higher. The error model includes a declared 5\% floor for the model's own imprecision, so the $\chi^2$ serves to rank the models and to locate their residuals. The residual structure is reported with its size. The forward $K^+$ fall below the data by \KplusIdPull{} times the identification systematic. Together with the forward $K^{*}(892)^0$ excess and the flat $\Lambda$ $x_F$ distribution, this residual points to how momentum is shared at the vertex where the leading baryon turns into a hyperon and emits a kaon. The forward $K^-$ show a separate residual in the shape of their $p_T$ spectrum. The $\Xi^-$ reaches \XiMinusRatio{} of the yield NA61/SHINE measured. The model locates this shortfall in the decay chains that start from strange baryons. The $\bar\Xi^+$ reaches \XiBarRatio{} of its measured yield. The very forward soft protons stand at \RegPSoftVfwd{} of the data. The charged multiplicity of a heavy excited system grows linearly with its mass, faster than the logarithmic growth measured in $e^+e^-$ annihilation. The observables that distinguish a cascade from string and fireball models lie in the fragmentation region. Wide-acceptance fixed-target spectrometers can measure them.
\end{abstract}

\section{Introduction}\label{sec:intro}

The NA49 experiment measured the double-differential cross sections of $\pi^\pm$, $K^\pm$, $p$, $\bar p$, and $n$ in p+p collisions at 158~GeV/$c$ beam momentum ($\sqrt{s} = 17.3$~GeV) over most of the forward hemisphere, with systematic uncertainties at the percent level~\cite{na49pppi2006,na49ppk2010,fischer2022piminus}. Fischer and co-workers interpret these spectra as the sum of two components, one per incoming proton, with a limited overlap region. They interpreted each component as the decay products of an excited hadronic system, that is, resonances cascading down from heavier objects~\cite{varga2003baryontransfer,chvala2004isospin,na49pcdisc2007,chvala2013backward,fischer2022piminus}. This line of analysis did not supply an event-level model of the picture. Such a model needs a definite set of exchange classes, excited systems with definite quantum numbers and masses, a decay rule, and exact conservation at every step. The spectra then follow from the model and can test it.

This paper develops that model and tests it. The model is named HEART, for Hadronic Excitations And Resonance Transitions. The name lists the two working parts of the event record, namely the excited hadronic systems the exchange produces and the successive two-body transitions by which they decay. The name also states the model's focus. It describes the soft heart of the strong force, the non-perturbative hadronic sector tabulated in the published NA49 cross-section data over $0.05 \le p_T \le 1.5$~GeV/$c$. A companion paper reviews the theoretical lineage of the two-component picture, its Regge organization, and its duality foundations~\cite{chvala2026theoryreview}. 

The HEART model has four layers that stay separate. The exchange sets the event class, the momentum transfer, and the masses of the two excited systems. The spectrum is a catalog of resonances below a duality boundary near 2~GeV and a resonant continuum with a declared density above it. The decay is a chain of quasi-two-body transitions through established hadrons. The hadron chemistry puts four flavor weights on the continuum vertices, beside the branching fractions of the Particle Data Group (PDG) for the catalogued states. The event record contains only hadrons, with no strings, diquarks, or partons. \NFitParamsCap{} parameters are fitted to the published NA49 cross-section data. Among them are the shares and direction laws of the two leading splits, the first decay steps that decide how the leading baryon or a leading meson leaves an excited proton (Section~\ref{sec:cascade}). The external inputs are the diffractive cross sections, the PDG tables, the pion-exchange flux, and the p+p elastic amplitude. The remaining discrete choices of the model, such as the exponent of the vertex slopes and the duality boundary, were set by paired comparisons (Section~\ref{sec:parameters}). In a paired comparison the alternatives are fitted to the same simulated events, so that their difference in $\chi^2$ is free of the sample's statistical fluctuation.

The comparison uses the data's own systematic uncertainties. The tabulated errors of the NA49 spectra are statistical. The papers give the systematic uncertainties separately, and these are correlated across the bins of a species. The fit carries them as systematic scales with the papers' priors. These scales are refitted, that is, profiled, at every evaluation of the $\chi^2$. A statement about a residual is then a statement against the data's stated uncertainty. The same error model is applied to three fixed-tune event generators, PYTHIA~8, EPOS-LHC, and UrQMD, on the same points.

The paper is arranged as follows. Section~\ref{sec:physics} states the physics picture and where HEART differs from the string and statistical generators. Section~\ref{sec:model} defines HEART layer by layer. Section~\ref{sec:comparison} defines the data, the estimator, the error model, and the fit. Section~\ref{sec:results} presents the fit, the spectra, the systematic scales and variants, the leading neutrons, and the transverse-momentum and multiplicity readouts. It continues with the generator comparison, the cross-checks outside the fit, and the structure that remains. Section~\ref{sec:discussion} discusses what separates a cascade from a string and from a fireball, and which measurements would decide between them. It names the spinless continuum as the model's largest approximation, with a route to a derived angular law. It sets the declared continuum density against the bootstrap condition. It also names the nuclear transfer as the next test. Section~\ref{sec:conclusions} concludes.

\section{Physics picture}\label{sec:physics}

\subsection{Two components and limiting fragmentation}

The inclusive distributions of the fragmentation regions approach energy-independent limits. This behavior is called limiting fragmentation~\cite{benecke1969limiting,feynman1969veryhigh}. The baryon-transfer analyses of Varga, Chvala, and co-workers, with baryon-free projectiles and tagged forward baryons in the NA49 data, showed that the target component of a collision does not depend on the projectile. The projectile component alone carries the baryon stopping, the shift of the incoming baryon toward central rapidity~\cite{varga2003baryontransfer,varga2004baryonpair,chvala2004isospin}. Fischer, Makariev, Varga, and Wenig built on this a global interpolation of the $\pi^-$ cross sections from $\sqrt{s} = 3$~GeV to 13~TeV. They argued that the soft sector is the decay of resonances cascading from heavier states~\cite{fischer2022piminus}. In a direct count of the $\pi^-$ from known resonance decays, they kept the two-body decays to final hadrons and the three-body decays of the $\eta$ and $\omega$. This count accounts for about three quarters of the inclusive $\pi^-$. They treat the result as a lower limit, with cascading and higher-multiplicity decays expected to fill the low-$p_T$ and low-$x_F$ region~\cite{fischer2022piminus}. Varga's decay simulation at Quark Matter 2006 uses the same two-body baseline with measured branching fractions. It assigns its low-$p_T$ low-$x_F$ shortfall to neglected many-body decays and cascading~\cite{varga2006pppa}. Rybicki's thesis simulates the hadronic two-body decay of the parent resonance at unit branching fraction in the studied channels~\cite{rybicki2003charged}. The present paper takes the two components as the two excited systems of one exchange and asks what else follows once their decay is specified.

\subsection{Regge exchange and the excited systems}

At $\sqrt{s} = 17.3$~GeV vacuum exchange dominates the total cross section. The subleading Reggeon exchanges fall with energy but remain visible in the charge and strangeness structure of the spectra~\cite{regge1959complex,donnachie1992totalxsec}. The inclusive fragmentation-region cross section is organized in the triple-Regge form~\cite{goulianos1983diffractive}. Single diffraction leaves one proton intact and excites the other into a system of mass $M$. Below the coherence limit $M^2/s \approx 0.1$ the mass follows $d\sigma/dM^2 \propto (M^2)^{-1-\epsilon}$, with $\epsilon = \alpha_{\mathbb{P}}(0) - 1$ the intercept of the Pomeron trajectory less one. Beside the triple-Pomeron term, pion exchange produces the leading neutrons. Its flux scales as $\xi^{1 - 2\alpha_\pi(t)}$, with $\alpha_\pi(t)$ the pion trajectory and $\xi \approx 1 - x_F$ for a neutron at $x_F$~\cite{goulianos1983diffractive,fischer2022piminus}. The hadronic state density rises steeply with mass~\cite{hagedorn1965thermo,frautschi1971bootstrap}, so an excited system above the resolved resonances is a dense spectrum of broad states. Its decay populates the final state through many intermediate hadrons.

The models compared in this paper differ in how they treat an excited hadronic system. The string generators compared here at their published tunes are PYTHIA, EPOS, and UrQMD~\cite{pythia83guide,eposlhc2015,urqmd4manual2025}. They treat an excitation heavier than about 2~GeV as a color string between a quark and a diquark. A universal fragmentation function hadronizes the string~\cite{andersson1983string}. EPOS identifies remnants below 2~GeV with named resonances~\cite{eposlhc2015}. UrQMD lists effective $N^*$ and $\Delta^*$ states to 4.8~GeV, which its dynamics excite only below its string threshold~\cite{urqmd4manual2025}. Statistical descriptions treat the excitation as a fireball~\cite{hagedorn1968ranft}. HEART treats an excitation of any mass below the onset of a hard scale as a hadronic resonance. Where resonances are resolved, the excitation is a catalogued state. Above the duality boundary it is a member of a resonant continuum with a declared spectral density. The excitation de-excites by successive two-body hadronic transitions. The identified-hadron observables that distinguish these treatments are the isospin and transverse momentum of the forward baryons, whether strangeness and baryon number are balanced locally, at the vertex that creates them, forward resonance signals with their mass tails and alignment, and the multiplicity of the excited system as a function of its mass (Section~\ref{sec:discussion}).

\section{The HEART model}\label{sec:model}

The companion paper gives the historical description of soft, non-perturbative hadron production in the fragmentation region, and the lineage of the model defined below~\cite{chvala2026theoryreview}. It follows limiting fragmentation from its statement by Benecke, Chou, Yang, and Yen and by Feynman, through the two-component picture of the spectra of the CERN Super Proton Synchrotron (SPS) by Fischer and co-workers, to the Regge organization of the fragmentation region and the diffractive mass law of the excited systems. Finite-energy sum rules and resonance duality motivate describing those systems by the resonance towers that build the exchange. The same paper records the cascade predecessors and names the fragmentation-region measurements that separate a resonance cascade from a string description.

This paper builds on that historical account without repeating it. What follows is the definition of the event-level realization and its fit to the published NA49 tables.

\subsection{Event classes and exchange}\label{sec:exchange}

An inelastic p+p event belongs to one of four classes. In non-diffractive production, both protons are excited. In single diffraction on the projectile or the target side, one proton line stays unexcited and the other is excited. In double diffraction both are excited at low mass. The single-diffraction classes hold three components, namely the triple-Pomeron coherence region, the leading-proton tail, and the one-pion exchange. The first two leave a proton intact. The charged branch of the third turns that proton into a leading neutron. The class fractions are external inputs, held fixed in the fit and anchored to the cross sections that follow. The single-diffraction fraction follows from $\sigma_{SD}(M^2/s < 0.05) = 4.9$~mb at $\sqrt{s} = 17$~GeV, the proton-dissociation measurement of Schamberger et~al.~\cite{schamberger1975}. The compilation of Goulianos and Montanha lists 4.87~mb at $\sqrt{s} = 16.2$~GeV, 4.96~mb at 17.6~GeV, and $4.94 \pm 0.08$~mb at 19.1~GeV. It assigns about 10\% to the normalization of these data~\cite{goulianos1999factorization}. The Fermilab and CERN Intersecting Storage Rings (ISR) measurements differ by 15\% to 20\% where they overlap. The systematic variants of the fit therefore move the value by $\pm 0.8$~mb (Table~\ref{tab:variants}). The measurement counts every proton that leaves the collision intact below $M^2/s = 0.05$. In the model these are the protons of the coherence component and of the neutral pion exchange, $p + p \to p + X^+$, which supplies \NeutralOPEWindowPct\% of them. The class's share of the inelastic cross section of the NA49 p+p analyses, $\sigma_{\rm inel} = 31.78$~mb~\cite{na49pppi2006}, is fixed so that the coherence component and the neutral exchange together reproduce the measured 4.9~mb below $M^2/s = 0.05$. The coherence component enters through the fraction of its mass law, Equation~\eqref{eq:ppp} integrated over the transfer, that lies below that edge on the class's mass window. The neutral exchange enters through the fraction of its flux below the edge. No measurement of the total double-diffraction cross section exists at this energy. It is estimated at $\sigma_{DD} = 1.5$~mb, between the extrapolations of the higher-energy data~\cite{goulianos2004wherestand}. The systematic variants move it to 0.5 and 2.5~mb. The non-diffractive cross section is the remainder to the total (Table~\ref{tab:classes}). The model has no central source, that is, no third excited system produced at central rapidity between the two proton lines. The central plateau, the flat region of the rapidity distribution around $y = 0$, is the overlap of the two cascades. A central color-singlet system with zero baryon number, charge, and strangeness, given a free fraction, fits to zero, so the model carries none.

\paragraph{Masses.} The diffractive systems of the coherence region, $M^2/s < 0.1$~\cite{goulianos1983diffractive}, follow the model's form of the triple-Pomeron term,
\begin{equation}
\frac{d^2N}{dM^2\,dt} = \frac{1}{\mathcal{N}}\,(M^2)^{-1-\epsilon}\;\frac{h_B(t)}{Z_B(M^2)}\;\left(\frac{M^2}{s}\right)^{-2\alpha'_{\mathbb{P}} t_{\rm eff}},
\qquad Z_B(M^2) = \int_{t_-(M)}^{t_+(M)} h_B(t)\,dt,
\label{eq:ppp}
\end{equation}
with the Donnachie--Landshoff trajectory $\alpha_{\mathbb{P}}(t) = 1 + \epsilon + \alpha'_{\mathbb{P}} t$, $\epsilon = 0.0808$ and $\alpha'_{\mathbb{P}} = 0.25$~GeV$^{-2}$~\cite{donnachie1992totalxsec}. Here $h_B(t)$ is the transfer law of the paragraph on momentum transfer below, $e^{Bt}$, continued for an intact proton beyond $|t| = 1$~GeV$^2$ with a flatter slope. $Z_B$ is its integral over the event's reachable transfer window $t_-(M) \le t \le t_+(M)$. $\mathcal{N}$ normalizes the density over the coherence region, and $t_{\rm eff} = \max(t, -1~{\rm GeV}^2)$. The last factor is the Pomeron's shrinkage, which narrows the transfer distribution as the rapidity gap $\ln(1/\xi)$ grows. Together with the power it reproduces the transfer dependence $\xi^{-1-\epsilon-2\alpha'_{\mathbb{P}} t}$ of the standard term, $\xi = M^2/s$. The division by $Z_B$ normalizes the vertex law at each mass. This is a phenomenological prescription of the model in place of a mass-independent triple-Regge residue. Without the shrinkage, $\alpha'_{\mathbb{P}} = 0$, the mass law integrated over $t$ is the power $(M^2)^{-1-\epsilon}$. At $t < 0$ the shrinkage factor raises the transfer slope by $2\alpha'_{\mathbb{P}}\ln(s/M^2)$, which is the shrinkage of the diffraction peak. The mass law integrated over $t$ is then the power times the ratio of the transfer integrals with and without the shrinkage. That ratio is smallest at small masses, where the rapidity gap is largest and the transfer distribution narrowest. It therefore flattens the mass law. The slope $\alpha'_{\mathbb{P}}$ acts for $|t| \le 1$~GeV$^2$, the range in which the linear trajectory is fitted. Beyond that range $t_{\rm eff}$ holds the exponent at its value at $t = -1$~GeV$^2$.

The same intercept enters the $\pi p$ cross section of the pion exchange below. Fits of the triple-Pomeron law to the diffraction data give a larger intercept, $\epsilon = 0.104$~\cite{goulianos1999factorization}. Fitted free, the spectra give $\epsilon = \EpsFree$, \EpsFreePull{} standard deviations below 0.0808. The model holds the Donnachie--Landshoff value, so one soft-Pomeron intercept serves the whole model.

The non-diffractive systems take their masses from $(M^2)^{-1-\epsilon_{\rm nd}}$ with $\epsilon_{\rm nd}$ fitted. Each line's mass starts at its own lowest open decay channel. The two lines' masses are drawn independently on the same window. A pair whose sum exceeds $\sqrt{s} - m_{\pi^0}$ is drawn again as a pair. Redrawing both members keeps the projectile and the target line statistically identical.

Beyond the coherence region a share $s_{\rm tail}$ of the single-diffraction class accounts for the leading-proton tail of the reggeized pion exchange beside the triple Pomeron,
\begin{equation}
\frac{d\sigma}{d\xi} \propto \frac{\xi}{1 + \exp[(\xi - 0.5)/0.1]}, \qquad 0.1 \le \xi \le 0.89,
\end{equation}
with $\xi = M^2/s \approx 1 - x_F$ for a proton at $x_F$. Its transfer follows $\exp(7t)$, and the proton stays intact.

\paragraph{One-pion exchange.} A second share $s_\pi$ of the single-diffraction class is the pion exchange proper, $p + p \to n + X^{++}$ and its isospin partner $p + p \to p + X^{+}$, in the squared-coupling ratio $2:1$ of the charged branch $p \to n\pi^+$ to the neutral branch $p \to p\pi^0$. The leading nucleon follows the pion flux of the fixed pion term of Appleby et~al.~\cite{appleby2016pomeron},
\begin{equation}
\frac{d^2\sigma}{d\xi\,dt} \propto \xi^{1 - 2\alpha_\pi(t)}\,\frac{-t}{(t - m_\pi^2)^2}\,F_1(t)^2\,\sigma_{\pi p}(\xi s)\,S^2(\xi, q),
\qquad \alpha_\pi(t) = 0.93\,(t - m_\pi^2),
\end{equation}
on $0.003 \le \xi \le 0.89$, with the proton's Dirac form factor
\begin{equation}
F_1(t) = \frac{4m_p^2 - 2.79\,t}{4m_p^2 - t}\;\frac{1}{(1 - t/0.71~\mathrm{GeV}^2)^2}
\end{equation}
and the Regge fit of the $\pi p$ total cross section, $\sigma_{\pi^+ p} = 13.63\,s'^{0.0808} + 27.56\,s'^{-0.4525}$~mb and the $\pi^0 p$ average with the coefficient 31.79~\cite{donnachie1992totalxsec}, with $s' \equiv \xi s$ the squared center-of-mass energy of the $\pi p$ sub-system.

The neutral branch of the exchange, which leaves the proton intact, uses the $\pi^0$ mass in its propagator and trajectory. Its fraction of the pion flux, $\zeta_0 = 0.349$, follows from integrating the charged and the neutral flux laws, with no free parameter. The exchange takes place inside the p+p collision, whose elastic scattering screens it. In the absorption model~\cite{sopkovich1962,gottfried1964absorption} the Born amplitude at impact parameter $b$ is multiplied by the elastic $S$-matrix $1 - \Gamma(b)$, with the p+p profile $\Gamma(b) = \sigma_{\rm tot}/(4\pi B)\,\exp(-b^2/2B)$ at the p+p total and elastic cross sections of the PDG compilation at this beam momentum, $\sigma_{\rm tot} = 38.75$~mb and $\sigma_{\rm el} = 6.97$~mb~\cite{pdg2022}. The slope $B = \sigma_{\rm tot}^2/(16\pi\sigma_{\rm el}) = 11.0$~GeV$^{-2}$ is the value at which this profile reproduces both cross sections. The measured slope of the forward diffraction peak at small $|t|$ is 11.4~GeV$^{-2}$ at this energy~\cite{bartenev1973slope}. The peak is curved, so its local slope at small $|t|$ exceeds the single slope of a Gaussian profile that carries the whole elastic cross section. The difference moves the absorbed flux by about 1\%.
 The flux's non-flip and flip helicity amplitudes, $\xi m_p F_1/(q^2 + \mu^2)$ and $q F_1/(q^2 + \mu^2)$, are transformed to impact parameter, screened, and transformed back, with $q$ the nucleon's transverse momentum and $\mu^2 = \xi^2 m_p^2 + (1 - \xi)\,m_\pi^2$. The survival factor $S^2(\xi, q)$ is their screened to unscreened ratio. In this prescription the pion-pole amplitudes are screened, and the Regge factor of the flux multiplies the result separately. The overall normalization of the exchange is not taken from this prescription. It remains the fitted share $s_\pi$. The pion's range $1/\mu$ shrinks with $\xi$. The exchanges that make heavy systems therefore act at small impact parameter, where the elastic scattering absorbs them most. $S^2$ averages 0.54 at $x_L = 0.9$, with $x_L$ the leading neutron's fraction of the beam momentum, 0.41 at $x_L = 0.75$, and 0.33 at $x_L = 0.6$. It falls from 0.9 to 0.16 between $p_T = 0.05$ and 0.6~GeV/$c$. The absorbed flux integrates to 0.49 of the unabsorbed one. The excited $\pi p$ system of mass $M$ hadronizes as a two-sided system, as a $\pi^+ p$ collision does. A baryonic line runs along its direction of flight and a $\pi$-like line runs opposite to it. Their masses are flat in $\ln M^2$, and the baryonic transfer follows $\exp(4t)$.

\paragraph{Momentum transfer.} Every two-body exchange takes its transfer from $\exp(Bt)$ with the factorized vertex slopes
\begin{equation}
B = \frac{b}{2}\left[\left(\frac{m_p}{M_1}\right)^{1.25} + \left(\frac{m_p}{M_2}\right)^{1.25}\right],
\end{equation}
which encodes the slope-mass correlation of diffraction dissociation, the flattening of the transfer distribution as the excited mass grows. Here $M_1$ and $M_2$ are the masses of the two excited systems, or $m_p$ for an intact proton, and $b$ is the fitted slope parameter. The exponent is fixed at 1.25. The values 1.0 and 1.5 cost \VertexExpLowCost{} and \VertexExpHighCost{} in $\chi^2$ on the same events. The intact proton of single diffraction receives its own vertex term $f$ in place of $(m_p/M_1)^{1.25}$. The separate term is needed because the vertex of an intact proton differs from the excited-vertex law evaluated at the proton mass. In the coherence-region events of single and double diffraction the shrinkage factor of Equation~\eqref{eq:ppp} raises the slope by $2\alpha'_{\mathbb{P}}L$ for $|t| \le 1$~GeV$^2$. Here $L = \max[0, \ln(s\,m_p^2/M_1^2M_2^2)]$ is the rapidity gap of the pair, which is $\ln(s/M^2)$ for an intact proton and zero when the two systems leave no gap. The masses of double diffraction follow the product of the two powers $(M_1^2)^{-1-\epsilon}(M_2^2)^{-1-\epsilon}$, with the factor $e^{2\alpha'_{\mathbb{P}} t_{\rm eff} L}$. Beyond $|t| = 1$~GeV$^2$ the intact proton's transfer continues with the slope $0.4B$. That slope reproduces the 2.5~GeV$^{-2}$ that the measured protons at $x_F = 0.95$ show beyond $|t| = 1$~GeV$^2$. The transfer kinematics are exact. The transfer is drawn only within the window that the two masses allow, and the event's four-momentum is conserved at every vertex.

\subsection{Spectrum of the excited systems}\label{sec:spectrum}

An excited system is a color-singlet hadronic state with definite baryon number, charge, and strangeness. Its spectrum has two parts. Above the duality boundary it is a member of a resonant continuum family of its quantum numbers, with the density
\begin{equation}
\rho(M) \propto \exp(M/T_H), \qquad T_H = 165~\mathrm{MeV},
\end{equation}
the Hagedorn growth of the state density with limiting temperature $T_H = 165$~MeV~\cite{hagedorn1965thermo,frautschi1971bootstrap}, on masses above the continuation floor of 2.0~GeV (Section~\ref{sec:cascade}). The families are the baryonic, the strange baryonic, the doubly strange baryonic, the neutral, charged, strange, and antistrange mesonic families, and the antibaryonic conjugates. A continuum state has a nominal width of 0.5~GeV and no spin or parity. Below the duality boundary the system is one of the catalogued resonances. There the resonances are resolved, and by local duality their sum stands for the exchange. A non-diffractive line of mass $M$ below \NdDualityMass~GeV, or a diffractive line below \DiffDualityMass~GeV, decays as one of the $N^*$ and $\Delta$ records of the PDG three-star and four-star tables at the line's own mass and momentum. The record $r$ is drawn with the weight $(2J_r + 1)\,\mathcal{B}_r(M)$. Here $\mathcal{B}_r$ is the normalized running-width spectral function the decayer samples for that record. The same line shape therefore describes a record wherever it appears, as an initial system or as a decay product. With running widths the $\Delta(1232)$ dominates the $N\pi$ spectrum near threshold, as measured. A fixed width would overstate the tail of the broad $N(1440)$ there. Below \RedrawMass~GeV, on the window between the $N\pi$ threshold and the $\Delta(1232)$ peak, the line's mass itself follows the mixture of these line shapes. The redraw maps the mass drawn from the continuum law onto the mixture by matching the two cumulative distributions, so it uses no additional random number. The principle of one line shape per record fixes the form of this redraw. The mass up to which it applies, its support, is set by the comparison with the data (Table~\ref{tab:fixed}). A support of 1.1~GeV nearly switches the redraw off and costs \RedrawOffCost{} in $\chi^2$. The spectra therefore need the redraw over a narrow window. In diffraction the vacuum exchange keeps the excited proton line at $I = 1/2$, so its low-mass record is drawn among the $I = 1/2$ records. The non-diffractive lines weigh their $I = 1/2$ and $3/2$ charge combinations equally.

Four mass scales act on the excited systems, namely
\begin{enumerate}
\item the duality boundary, \NdDualityMass~GeV on a non-diffractive line and \DiffDualityMass~GeV on a diffractive line, below which an initial system is one of the catalogued records;
\item the redraw support, \RedrawMass~GeV, below which the line's mass follows the mixture of the records' line shapes;
\item the continuation floor, \ContinuationFloor~GeV, the lowest mass of a continuum state and of a continuation daughter; and
\item the continuation opening near 2.3~GeV, where a pion sibling leaves a daughter at the floor with the minimum step momentum (Section~\ref{sec:cascade}).
\end{enumerate}
The duality boundary decides which description an initial system receives. The floor decides where the continuum ends. The two thresholds therefore answer different questions. The first concerns the start of a line, and the second the end of its cascade. The comparison with the data sets them separately (Table~\ref{tab:fixed}). A continuum state below the continuation opening is a floor state, which ends in one direct pair.

\subsection{Decay cascade}\label{sec:cascade}

A continuum state de-excites in steps, and each step makes two decisions. A state has two kinds of open channels. Its direct pairs are the isospin-complete two-body channels of its family into established hadrons. In its continuation channels one established sibling is emitted, and the remaining state continues as a lighter member of a continuum family with the matching quantum numbers. The first decision selects between the two pools. A state continues while any continuation channel is open, and it ends in a direct pair when none is. A continuation channel opens when the sibling's decay momentum at the daughter's lowest mass reaches 0.25~GeV/$c$. The second decision draws one channel of the pool with the weight
\begin{equation}
w_c(H) \propto C_{\rm iso}\,C_{\rm spin}\,q^*_c\;\gamma_s^{\,n_{s\bar s}}\;r_V^{\,n_V}\;r_{B\bar B}^{\,n_{B\bar B}}\;a_{YK}^{\,n_{YK}},
\label{eq:weights}
\end{equation}
normalized over the pool. Its factors are the isospin Clebsch--Gordan factor $C_{\rm iso}$, the spin counting $C_{\rm spin}$ (the product of $2J + 1$ over the established daughters), the two-body decay momentum $q^*_c$ of the channel, and the four flavor factors of Section~\ref{sec:chemistry}. A direct pair takes $q^*$ at the pole masses of its two hadrons. A continuation channel takes $q^*$ at an effective daughter mass, $m_{\rm eff} = \max(2.0~\mathrm{GeV},\, M - m_{\rm sib} - 0.5~\mathrm{GeV})$. That mass is the upper edge of the daughter's window, lowered by a fixed released energy of 0.5~GeV. Where the baryon-resonance pairs join a pool, its weights are rescaled so that the baryon-pair share of the pool keeps its value without them. The factor $r_{B\bar B}$ thereby sets the same baryon-pair share in every pool, whether or not the pool includes baryon-resonance pairs. Once a continuation channel is drawn, the daughter's mass is drawn on $[2.0~\mathrm{GeV},\, M - m_{\rm sib}]$ from the family's density times the two-body decay momentum with the $P$-wave barrier,
\begin{equation}
p(m) \propto \exp(m/T_H)\; q^{*3}\,F_1(q^*)^2, \qquad q^* = q^*(M; m, m_{\rm sib}),
\end{equation}
with $q^*$ the two-body decay momentum in the rest frame of the parent state and $F_1$ the Blatt--Weisskopf factor at interaction radius $r_0 = 1$~GeV$^{-1}$. The step that creates a baryon--antibaryon pair from a mesonic state draws the daughter's mass with an $S$-wave factor, $L = 0$, in place of the $P$ wave. The continuum states carry no spin or parity, so this barrier is an assumed sampling law. It shapes the daughter's mass distribution and leaves the decay angles untouched. For a floor state of the charged and strange mesonic families, the pool of direct pairs also holds heavier pairs ($\rho\omega$, $\rho\rho$, $f_2\pi$, $a_2\pi$, $a_1\pi$, $b_1\pi$, $\rho(1450)\pi$, $\phi\pi$, $K^*\bar K^*$; $K^*\rho$, $K^*\omega$, $K\eta$, $\phi K$, $K_1\pi$, $K_2^*\pi$, $K f_2$, and their isospin partners), which join the light pairs with the same weight law. The direct pairs of a baryonic line are the nucleon and the $\Delta$ with the pseudoscalar, vector, and scalar mesons and the axial nonets, and the hyperons and hyperon resonances with the kaons, the kaon vectors, and the excited kaons (associated production). The direct pairs of a mesonic line are the mesons among themselves and, above threshold, a baryon with an antibaryon. These include the nucleon resonances with their antibaryons at matched charge. The neutral mesonic family also offers every pair of octet hyperons and antihyperons its quantum numbers allow ($\Lambda\bar\Lambda$, $\Sigma\bar\Sigma$, $\Xi\bar\Xi$, and the mixed $\Lambda\bar\Sigma^0$ and $\Sigma^0\bar\Lambda$), at the same weight law as its nucleon pairs. A strange baryonic line pairs the $\Xi$ and the $\Xi(1530)$ with a kaon and continues into the doubly strange continuum, an isospin doublet that decays through $(\Xi, \pi)$, $(\Xi(1530), \pi)$, $(\Lambda, \bar K)$, $(\Sigma, \bar K)$, $(\Sigma(1385), \bar K)$, and $(\Omega^-, K)$.

The decay angles are isotropic in the state's rest frame, with three exceptions that the leading particles require. The first two act at the baryonic root of a cascade, in a fixed order. In the first exception, at \MesonSplitPct\% of the eligible roots an established meson leaves along the incoming baryon, the leading-meson split. In the second, at \BaryonSplitPct\% of the remaining roots the baryon leaves beside a mesonic remainder, the leading-baryon split. Where both splits are open, the meson branch therefore takes \MesonSplitPct\% of the roots, the baryon branch $\MesonSplitRemainder \times \BaryonSplitShare = \BaryonBranchPct\%$, and the ordinary draw takes the remaining \OrdinaryBranchPct\%. A split whose channels are closed by threshold drops out. These are the probabilities of the two aligned branches. The ordinary draw contains baryon and meson channels as well. In the baryon branch the channel is drawn among the split channels, each an established baryon beside a mesonic continuum state. The remainder's mass $\mu$ is flat in $\ln\mu^2$ on its window. The baryon leaves along the incoming baryon with $dP/d\cos\theta \propto \exp(\kappa_B\cos\theta)$, the form the flat $x_F$ distribution of the leading protons requires. In the meson branch the meson leaves along the incoming baryon with $\kappa = 2b_M q^{*2}$, the source of the very forward $\pi^-$. In both branches $\theta$ is the angle to the incoming baryon's direction in the root's rest frame, the Gottfried--Jackson axis of the exchange. Third, below the root, when a baryonic state continues into a baryonic continuum daughter beside a non-strange hadron, it sends that daughter along the line's beam axis with $dP/d\cos\theta \propto \exp(\kappa_L\cos\theta)$, $\kappa_L = \AlignKappa$. The line thereby keeps its baryon leading, and the emitted hadron recoils toward the center. With isotropic steps a line shares its momentum among its decay products roughly in proportion to their masses, so its final proton keeps a small part of it. The steps that transfer strangeness keep the isotropic decay. The strength $\kappa_L$ is a fixed choice, set by paired comparisons (Table~\ref{tab:fixed}). Against isotropic continuing steps it lowers $\chi^2$ by \AlignGain{}, in the protons and the pions together. The two shares $F_B$ and $F_M$ and the two direction laws $\kappa_B$ and $b_M$ are fitted parameters (Table~\ref{tab:parameters}), at $F_M = \ValFM$, $F_B = \BaryonSplitShare$, $\kappa_B = \SplitKappa$, and $b_M = \MesonSplitSlope$~GeV$^{-2}$. The event record stores each root's branch, its direction, and its decay momentum. The fit therefore moves the four parameters by exact reweighting of the generated events, with no new sample. The antibaryonic families are the conjugates of the baryonic ones, record for record. A baryon--antibaryon pair created at a mesonic vertex continues through the baryonic or the antibaryonic family, each with half the weight.

Every established hadron decays through its PDG table, namely the $N^*$ and $\Delta$ at every hadronic mode of the 2025 tables, the $\rho$, $\omega$, $\phi$, $f_2$, $a_2$, $K_1(1270)$, and $K_2^*(1430)$ at every mode of at least 1\% of the 2024 tables with the four-body and missing-state modes through their three-body final states, and the $\omega$, $\eta$, and $\eta'$ three-body decays at their measured Dalitz distributions. The same tables decay the hyperons from $\Lambda(1405)$ to $\Sigma(1915)$, the $\Xi$ and $\Xi(1530)$, the light scalars $f_0(500)$, $f_0(980)$, and $a_0(980)$, the axial nonets $a_1$, $b_1$, $f_1(1285)$, $h_1(1170)$, and $K_1$, and the excited kaons $K^*(1410)$, $K_0^*(1430)$, $K_1(1400)$, and $K^*(1680)$. Every two-body charge state carries the exact Clebsch--Gordan factor of its isospin multiplet. Three-body decays follow flat phase space where no Dalitz measurement exists. Broad siblings take their masses from their running-width line shapes inside the available window. Two two-body modes are closed at the pole and open above it, namely $f_0(980) \to K^0\bar K^0$ and $a_0(980) \to K\bar K$. The branching fraction of each mode is evaluated half a width above the larger of the pole mass and the threshold. This gives the two modes 0.058 and 0.066 of the decays. The radiative modes of the baryon resonances, 0.6\% of the $\Delta(1232)$ decays and less for the heavier states, are left out. They change the pion yields by less than 0.1\%. The cascade ends at the weakly decaying hadrons and at the $\Sigma^0$. The model's observables then perform those weak and electromagnetic decays whose products the NA49 tables include. The NA49 pion tables are corrected for the decay products of $K^0_S$, $\Lambda$, $\Sigma^\pm$, and $\bar\Lambda$~\cite{na49pppi2006}. The proton and antiproton tables are corrected for those of $\Lambda$, $\Sigma$, and their antiparticles~\cite{na49ppproton2010}. The kaon tables carry no such correction, since the only weak decay that feeds them, that of the $\Omega^-$, is negligible~\cite{na49ppk2010}. Strong and electromagnetic decays stay in all the tables by definition. The observables therefore decay the $\Sigma^0$ to $\Lambda\gamma$, the $\Xi$ to $\Lambda\pi$, and the $\Omega^-$ through $\Lambda K^-$, $\Xi^0\pi^-$, and $\Xi^-\pi^0$. The antiparticles decay through the charge-conjugate modes. The $K^0_S$, $K^0_L$, $\Lambda$, and $\Sigma^\pm$ stay final-state hadrons. The $\Xi$ decay pions add less than 0.1\% to the pion yields.

\subsection{Hadron chemistry}\label{sec:chemistry}

The four flavor factors of Equation~\eqref{eq:weights} act on the continuum vertices only. A catalogued parent keeps its measured branching fractions. The exponent of $\gamma_s$ counts the strange quark--antiquark pairs created at the vertex, $(\sum_i |S_i| - |S_H|)/2$ over the daughters. It adds the hidden strangeness of daughters without net strangeness, one pair for the $\phi$ under ideal mixing and 0.6 for the $\eta'$, its quark-model $s\bar s$ content. The exponent of $r_V$ counts the established vector mesons among the daughters ($J = 1$, negative parity, namely $\rho$, $\omega$, $\phi$, and $K^*$ with their excitations). The exponent of $r_{B\bar B}$ counts the baryon pairs created at a continuing step, $(\sum_i |B_i| - |B_H|)/2$. The direct $p\bar p$ and $n\bar n$ pairs of a floor state carry no $r_{B\bar B}$ factor. Their bare two-body weight gives them 0.04 to 0.05 of the decays of a neutral state at 2.0 to 2.3~GeV. With these weights the inclusive antiproton yield of the fitted model is \AntiprotonYieldRatio{} of the measured one (Table~\ref{tab:segments}). The factor $a_{YK}$ applies once to a channel of a non-strange baryonic parent that produces a hyperon together with a kaon. The factors are effective hadronic channel weights. Each equals one when the channel weights follow isospin, spin counting, and phase space alone. The fit finds $\gamma_s$ \GammaSRelation, $r_V$ \RVRelation, $r_{B\bar B}$ \RBBRelation, and $a_{YK}$ \AYKRelation.

\subsection{Conservation and reproducibility}\label{sec:conservation}

Every vertex of the event record conserves four-momentum exactly, and baryon number, charge, and strangeness as integers. Every channel opens only above its threshold. A self-check suite verifies both conditions on every registered decay mode and on samples of generated events. Given its random seeds, the generator reproduces every event exactly. Every vertex draws a fixed budget of uniform numbers from named substreams, namely the topology selection, the spectral state, the exchange kinematics, the decay channel, and the decay kinematics. An event is therefore reproducible bit for bit. The named streams also allow two model variants to be compared on the same random numbers, which cancels most of the statistical noise in their difference. A change that alters an event's decay history shifts which random numbers its later steps receive, so the cancellation is incomplete. A paired difference is therefore quoted with an uncertainty measured from independent replicas (Section~\ref{sec:fit}).

\subsection{Parameters and fixed choices}\label{sec:parameters}

\NFitParamsCap{} parameters are fitted, namely the non-diffractive mass law $\epsilon_{\rm nd}$, the vertex slope $b$, the intact proton's vertex term $f$, the tail share $s_{\rm tail}$ and the one-pion-exchange share $s_\pi$ of single diffraction, the four flavor weights, the shares $F_B$ and $F_M$ of the leading-baryon and leading-meson splits with the baryon split's direction $\kappa_B$ and the meson split's transverse slope $b_M$, and the isovector exchange fraction $f_{\rm iv}$.

The minimum in $f_{\rm iv}$ lies on its boundary at zero. The profile of $\chi^2$ on the allowed side rises from that boundary with a slope of \FivSlope{} per unit of $f_{\rm iv}$ and a positive curvature. The fit quotes the one-sided limit $f_{\rm iv} < \FivLimit$ at 95\% confidence, at the point where the refitted profile crosses $\Delta\chi^2 = 2.71\,S^2$ above the $f_{\rm iv} = 0$ minimum. The limit holds within the model as specified. It is conditional on the fixed model choices, the discrepancy floor, and the scaling by $S$.

The other inputs are fixed, and none is varied in the fit (Table~\ref{tab:fixed}). The diffractive $\epsilon$ and $\alpha'_{\mathbb{P}}$, the coherence edge, the continuum density, and the class anchors come from the literature and the measurements stated above. The discrete choices of the cascade are the vertex exponent, the duality boundaries, the redraw support, the continuation floor, and the step minimum. Each of them changes which decay chain an event receives, or enters the transfer normalization of every class. Reweighting one generated sample therefore cannot vary them, and each alternative needs its own sample. They are set by paired comparisons. Table~\ref{tab:fixed} gives each with the $\chi^2$ cost of its neighboring values. The costs of the neighboring values differ widely between the choices. The continuation floor, the step minimum, and the vertex exponent are fixed firmly. The non-diffractive duality boundary and the redraw support are determined more weakly. The neighboring values of the diffractive duality boundary differ by less than the statistical noise. The quoted errors of the \NFitParams{} parameters are conditional on these discrete choices. The systematic variants of Table~\ref{tab:variants} move the discrepancy floor, the common normalization, and the two diffractive cross sections. They leave these choices in place.

\begin{table}[htbp]
\centering
\footnotesize
\begin{tabular}{p{3.3cm}lp{2.7cm}p{5.6cm}}
\toprule
Fixed choice & Value & Set by & Neighboring values: $\Delta\chi^2$ \\
\midrule
\FixedChoiceRows
\bottomrule
\end{tabular}
\caption{The fixed choices of the model. The upper rows come from the literature and the measurements. The systematic variants of Table~\ref{tab:variants} move the two diffractive cross sections, and the diffractive intercept is also fitted free under the same diffractive law. The lower rows are set by the comparison with the data. For these, $\Delta\chi^2$ is the change of the six-species $\chi^2$ against the adopted value when one choice takes the neighboring value, with the species normalizations refitted and the \NFitParams{} parameters held. Each comparison uses the same simulated events, two replica blocks of $10^6$ events at a fitted point of the model, and its error is the paired jackknife error of the two blocks.}
\label{tab:fixed}
\end{table}

\section{Comparison method}\label{sec:comparison}

\subsection{NA49 data}

The fit uses the NA49 double-differential invariant cross sections $f(x_F, p_T) = E\,d^3\sigma/dp^3$ of $\pi^-$ and $\pi^+$~\cite{na49pppi2006}, $K^+$ and $K^-$~\cite{na49ppk2010}, and $p$ and $\bar p$~\cite{na49ppproton2010}, in the tables of the SPS HADRONS collection~\cite{spshadrons2026}. They cover $0 \le x_F \le 0.95$ and $0.05 \le p_T \le 1.5$~GeV/$c$ with 281, 260, 153, 154, 284, and 122 points, 1,254 in all. The eight points of the neutron $dn/dx_F$ at $x_F = 0.1$ to $0.9$ join them. The $\pi^-$ table also lists an entry at $x_F = 0.85$, $p_T = 1.5$~GeV/$c$ with value and error zero. It is not a measurement, carries no weight in the fit, and is not counted. The tables are normalized to the inelastic cross section, $\sigma_{\rm inel} = 31.78$~mb. The comparison is made in $f/\sigma_{\rm inel}$, the cross section per inelastic event.

\subsection{Estimator}

The model's prediction at a tabulated point is the average of $f/\sigma_{\rm inel}$ over the point's bin, whose edges are the midpoints to the neighboring tabulated values,
\begin{equation}
\hat f_{ij} = \frac{c_{ij}}{\pi\sqrt{s}\,N_{\rm ev}\,\Delta x_{F,i}\,\Delta p_{T,j}}\sum_{k \in (i,j)} \frac{E_k}{p_{T,k}},
\end{equation}
where $N_{\rm ev}$ is the number of generated inelastic events, $\Delta x_{F,i}$ and $\Delta p_{T,j}$ are the bin widths, and the sum runs over the tracks $k$ of the species in bin $(i,j)$ with center-of-mass energy $E_k$ and transverse momentum $p_{T,k}$ at $\sqrt{s} = 17.270$~GeV. The data are quoted at points, while the model is counted in bins. The factor $c_{ij}$ converts the bin average into the value at the tabulated point. It is computed from the data's own local shape. Along $x_F$ at fixed $p_T$, and along $p_T$ at fixed $x_F$, $\log f$ is fitted through the point and its nearest tabulated neighbors, quadratically with two neighbors and linearly with one. The ratio of the value at the point to its average over the bin is the product of the two one-dimensional ratios. The factors depend on the data alone and are fixed per bin. Their median is \BCDataMedian. They depart from 1 by a median of \BCDataDepartPct\%, by more than 5\% at \BCDataOverFive{} of the 1,254 points, and by at most \BCDataMaxPct\%. The largest factors lie at $p_T = 0.9$~GeV/$c$ in the kaon and baryon tables, where the tabulated step widens from 0.1 to 0.2~GeV/$c$ and the bin extends further above the point than below it. The factors correct the model by the data's local curvature. Computed instead from the model's own local shape, they change by a median of \BCModelChangePct\%. They change by more than 5\% at \BCModelOverFive{} points, \BCModelBaryons{} of them in the proton and antiproton tables, mostly at the edges of the measured region where the model's statistical precision is lowest. At the fitted point, with the scales held, that change raises the $\chi^2$ of the six spectra by \BCModelChi. The neutron $dn/dx_F$ is the model's forward-hemisphere neutron count per inelastic event in the table's bins. The total yields are the model's full-phase-space counts per inelastic event, compared with the published total multiplicities.

\subsection{Error model}\label{sec:errors}

The tabulated errors of the NA49 spectra are statistical. The papers give the systematic uncertainties separately, namely 2.0\% for the pions~\cite{na49pppi2006}, 2.5\% for the protons, and 3.3\% for the antiprotons~\cite{na49ppproton2010}. For the pions that total combines a normalization component of 1.5\% with feed-down and acceptance components of 0.5\% to 1.5\% and smaller terms~\cite{na49pppi2006}. The kaons carry 2.2\% at $x_F \le 0.2$. At $x_F \ge 0.25$ they carry a particle-identification uncertainty that rises with $x_F$, from about 4\% to 12\% for $K^+$ and from 0 to 6\% for $K^-$, as the proton fraction overwhelms the kaon in the fits of the specific energy loss $dE/dx$ in the time projection chambers (TPCs)~\cite{na49ppk2010}. The neutron table's errors of 10\% to 28\% per point represent the systematic uncertainties of the calorimeter's energy scale and unfolding, correlated across $x_F$. These uncertainties are correlated across the bins of a species and enter as systematic scales. A systematic scale multiplies the model's prediction for one species by a factor the papers' systematic uncertainties allow. The comparison then tests the shape of the spectra within the normalization the papers allow. For each point $i$ the model is compared as
\begin{equation}
\chi^2 = \sum_i \frac{\left[d_i - m_i\,(1 + \sum_k A_{ik}\eta_k)\right]^2}{\sigma_i^2 + (0.05\,d_i)^2 + s_i^2} + \sum_k \left(\frac{\eta_k}{p_k}\right)^2.
\end{equation}
Here $d_i$ is the measured cross-section value, $m_i$ the model prediction, $\sigma_i$ the tabulated statistical error, and $s_i$ the Monte Carlo error of the prediction. The 5\% floor is the model's own allowance for a shape it does not describe. The Monte Carlo error is the spread of the unweighted replica blocks at the trial class fractions. It approximates the error of the reweighted prediction. The approximation is closest near the generation point, and regeneration keeps the minimum near that point (paragraph on regeneration below). At the minimum the spread of the reweighted blocks exceeds it by a median of \MCSpreadMedianPct\% (by \MCSpreadRangePct\% across the species). The Monte Carlo error is a median \MCOverSigma{} of the statistical error. With the reweighted spread in its place, the $\chi^2$ minimum is lower by \MCDeltaChi, which changes $S$ and every quoted error by \MCSChangePct\%. The scales $\eta_k$ act through their amplitudes $A_{ik}$. Each species has one normalization with the prior $p_k = 1.5\%$, the pion paper's normalization component is applied to every species. Each kaon charge has one identification scale above $x_F = 0.25$, with the amplitude the kaon paper states and prior 1 in units of that amplitude. That amplitude is 4\% at $x_F = 0.25$, rising linearly to 12\% at 0.40 for $K^+$, and 0 to 6\% at 0.50 for $K^-$. The neutron table has one scale with the prior 10\%. Each neutron point keeps as its own error the part of its tabulated error that the common scale does not cover, at least 5\% of the point. The separate 5\% model floor is added to it. The scales enter the prediction linearly, so their best values follow in closed form at every evaluation of the $\chi^2$. Their pulls measure the model's residuals against the data's stated systematic uncertainties. The feed-down and acceptance components of the quoted totals are not modeled as separate scales. They are smaller than the 5\% floor, which covers them. That floor remains the model's allowance for its own shape residuals. Table~\ref{tab:errormap} maps each quoted component to its term in the comparison.

\begin{table}[htbp]
\centering
\footnotesize
\setlength{\tabcolsep}{4pt}
\begin{tabular}{@{}lll@{}}
\toprule
Quoted component & Size & Term in the comparison \\
\midrule
Normalization, each species & 1.5\% (pions) & scale $\eta_k$ per species, prior 1.5\% \\
Feed-down, acceptance, other & 0.5\%--1.5\% (pions) & not modeled; inside the 5\% floor \\
The remainder of the $p$, $\bar p$ totals & to 2.5\%, 3.3\% & not modeled; inside the 5\% floor \\
Remainder of the kaon total, $x_F \le 0.2$ & to 2.2\% & not modeled; inside the 5\% floor \\
Kaon identification, $x_F \ge 0.25$ & 4\%--12\% ($K^+$), 0--6\% ($K^-$) & one scale per charge, amplitude as stated \\
Neutron energy scale and unfolding & 10\%--28\% per point & common scale, prior 10\%, \\
 & & and the point remainder, at least 5\% \\
Normalization common to all species & not quoted & variants, common shift $\pm 2\%$ \\
Model shape & --- & 5\% floor on every point \\
\bottomrule
\end{tabular}
\caption{The systematic uncertainties quoted by the NA49 p+p papers~\cite{na49pppi2006,na49ppproton2010,na49ppk2010} and the term of the comparison that carries each. The species normalization scales are independent of each other. The common shift of the variants multiplies every prediction of the model and leaves the priors of the scales unchanged. The neutron point remainder and the model floor are two separate 5\% terms.}
\label{tab:errormap}
\end{table}

The $\chi^2$ of this comparison is to be read with its floor in mind. The statistical errors of the pion tables are 2\% to 5\% per point, so without the floor a 5\% shape residual costs several units per point. The floor, therefore, sets the scale of every $\chi^2$ quoted below. The floor variants of Table~\ref{tab:variants} move the total by thousands. They move the parameters by at most \VarFloorMaxMove{} errors. The claims rest on three things the floor does not change, namely the ranking of the models on one objective, the localization of the residuals in the $(x_F, p_T)$ plane, and the pulls of the profiled scales against the papers' stated systematics. The value $\chi^2/{\rm ndf} = \FitSsq{}$ describes the fit under this floor. It is not a goodness-of-fit test of the model.

\subsection{Fit procedure}\label{sec:fit}

\paragraph{Generation and reweighting.} The model is generated once at a stated parameter point with random seeds fixed in advance, in 32 replica blocks of $10^5$ events per class ($12.8 \times 10^6$ events in all). Each event records the draws of its exchange and of every cascade vertex. Every other parameter point is evaluated on this sample by importance reweighting. An event's weight is the product, over its exchange and its continuum vertices, of the probability of its recorded draws at the new point divided by the probability at the generation point. The factors in that ratio are the mass laws for $\epsilon_{\rm nd}$, the transfer densities for $b$ and $f$, the class shares for $s_{\rm tail}$ and $s_\pi$, and the normalized channel weights of Equation~\eqref{eq:weights} for the flavor factors at each vertex, recorded for every vertex. The objective is thus a smooth function of the parameters on one sample. Two parameter points are compared with the same events. The effective sample fraction of each class is monitored. It is the effective number of events of the weighted sample, $(\sum w)^2/\sum w^2$, divided by the number generated. The fit is accepted only when every class stays above 0.1. At the minimum reported here the four classes lie between \ESSMin{} and \ESSMax{}. The scales of the error model are profiled in closed form inside the objective at every evaluation.

\paragraph{Minimization and errors.} The minimum is found from dispersed starts by a bounded minimization on transformed coordinates, logarithmic for the slopes and weights, and logistic for the shares. The Hessian at the minimum is taken with steps fixed in advance. The errors are its inverse, scaled by $S = \sqrt{\chi^2/{\rm ndf}}$. The isovector exchange is tested by a second minimization with $f_{\rm iv} = 0$. The free minimum lies on that boundary, so the two minima coincide ($\Delta\chi^2 = 0$). The profile of $f_{\rm iv}$ refits the other \NFitParamsOther{} parameters at $f_{\rm iv} = 0.01$, 0.02, 0.03, and 0.045 and gives $\Delta\chi^2 = \FivDeltaOne$, \FivDeltaTwo, \FivDeltaThree, and \FivDeltaFour. A curve $\Delta\chi^2 = a f_{\rm iv} + c f_{\rm iv}^2$ through the boundary describes these points within \FivFitResid. Its crossing of the threshold gives the limit. The Hessian includes the boundary direction of $f_{\rm iv}$. The errors of the other \NFitParamsOther{} parameters are read from its inverse, which treats that direction as open on both sides. They are therefore at least as large as with $f_{\rm iv}$ held at zero. The 32 blocks are also fitted separately as a replica test, and each block's optimum is compared with the expected spread. The fit is recorded only when every block's optimum lies within that spread.

\paragraph{Regeneration and resolution.} The fit reweights one generated sample to every trial point. Within a fit, the model is regenerated when the effective sample fraction of any class falls below 0.3. The reweighting is exact in expectation, but far from the generation point, the reweighted sample fluctuates more. The model is therefore regenerated at the fit's optimum with new random seeds and refitted until the optimum stops moving in one direction. The fit reported here was generated at the optimum of the preceding regeneration, and its parameters lie within \RegenMaxMove{} errors of that point. Independent samples generated at one point differ by up to \RegenReplicaMaxDiff{} errors, in the meson-split slope $b_M$ and in $r_V$, the parameters with the largest remaining shifts. Independent samples of the model at or next to this optimum give $\chi^2$ values with a standard deviation of \RegenScatter{} (\RegenSamples{} samples). That spread is the resolution of the comparison. Two independent fits of one model differ by about \RegenPairSpread{} in $\chi^2$. A component of the model is kept or set aside against that scale.

The components of Section~\ref{sec:model} entered one at a time, each as a refit against the model without it. A component was adopted when its physics is established and the data do not reject it beyond the resolution. Where two forms lie within the resolution, the more complete and more parsimonious one is kept. Every alternative set aside is recorded with its cost. The paired comparisons quoted in Section~\ref{sec:model} evaluate a changed model on the reference sample's own seeds, in two blocks of $10^6$ events each, with the per-species normalizations fitted. Their resolution is $\pm 20$ to $\pm 100$ in $\chi^2$, depending on how much of the event history the change preserves.

\paragraph{Advance specification and result record.} The specification of each fit is written down and checksummed before its events are generated. It includes the version of the model's code, the parameters and their starting point, the random seeds, the external inputs, the error model, and the readouts. The output directory carries a manifest with the checksum of every file. Every event is reproducible from its named random streams (Section~\ref{sec:conservation}). The systematic variants of the comparison are refits of the same sample with the discrepancy floor at 3\%, 7\%, and 10\%, a common shift of $\pm 2\%$ in the normalization of every species' prediction, $\sigma_{SD}(M^2/s < 0.05)$ at 4.1 and 5.7~mb, and $\sigma_{DD}$ at 0.5 and 2.5~mb. One fit, from generation to the final record, takes about four hours of wall time on a small cluster.

\subsection{External generators}\label{sec:generators}

PYTHIA~8 (SoftQCD, default tune)~\cite{pythia83guide}, EPOS-LHC through the Cosmic Ray Monte Carlo interface (CRMC)~\cite{eposlhc2015}, and UrQMD~4~\cite{urqmd4manual2025} were run at 158~GeV/$c$ with $\GeneratorEvents \times 10^6$ inelastic events each. Their final-state tracks were binned on the same tabulated points by one estimator and scored under the same error model, including its profiled scales. No generator was retuned on these spectra. The comparison tests each generator at its published tune. It makes no claim that the models have equal numbers of adjustable parameters. Its limits are as follows. PYTHIA~8's default (Monash) tune is a collider tune, and its authors did not optimize its SoftQCD description at $\sqrt{s} = 17.3$~GeV. EPOS-LHC descends from model versions tuned with fixed-target data, so of the three generators it has the strongest claim to describe this energy. UrQMD's hadronic cascade applies at this energy, but it hands excitations above its string threshold to a FRITIOF string. SMASH~3.3~\cite{smash2016} was run at its released defaults and left out, because its string channels do not engage for nucleon--nucleon collisions at this energy. Its events therefore hold no antibaryons and a strongly suppressed $K^-$ yield. A like-for-like comparison would need settings beyond the published ones. Readers can extend the comparison themselves. The SPS-HEART repository listed under Code availability includes an event generator for the fitted model, so HEART can be tested against other data or against another model.

\section{Results}\label{sec:results}

\subsection{Fit quality and fitted values}\label{sec:fitresult}

The fit gives $\chi^2 = \FitChiSq$ on \FitNdf{} degrees of freedom, $S^2 = \FitSsq$. The degrees of freedom are the 1,262 measured points less the \NFitParams{} parameters. The nine profiled scales are not subtracted from the degrees of freedom, because each carries a Gaussian prior penalty inside the $\chi^2$. On average the residuals are \FitS{} times the uncertainties of the error model of Section~\ref{sec:errors}. The 5\% floor sets the scale of those uncertainties. The Hessian is positive definite, and every replica block's optimum lies within the expected spread. The smallest effective sample fraction is \ESSMin{}. Table~\ref{tab:parameters} lists the parameters with their errors scaled by $S$. Table~\ref{tab:classes} lists the event classes.

\begin{table}[htbp]
\centering
\setlength{\tabcolsep}{4pt}
\begin{tabular}{@{}lp{0.545\textwidth}c@{}}
\toprule
Parameter & Meaning & Fitted value \\
\midrule
\ParTableRows
\bottomrule
\end{tabular}
\caption{The \NFitParams{} fitted parameters of the model at the minimum, with errors scaled by $S = \FitS$. The isovector exchange fraction lies on its boundary at zero. Its entry is the one-sided 95\% limit of its profile. The model has no central source.}
\label{tab:parameters}
\end{table}

\begin{table}[htbp]
\centering
\begin{tabular}{lcc}
\toprule
Class & Fraction of $\sigma_{\rm inel}$ & $\sigma$ [mb] \\
\midrule
\ClassRows
\bottomrule
\end{tabular}
\caption{The event classes at the fitted point. Single diffraction totals \SDTotalMb{}~mb. Of that, \SDWindowMb{}~mb are intact protons below $M^2/s = 0.05$, the measured value, and \NeutralOPEWindowPct\% of them come from the neutral pion exchange. Per side the leading-proton tail accounts for \TailMbPerSide{}~mb. The one-pion exchange accounts for \OPEMbPerSide{}~mb per side, \OPEOverUnabsorbed{} of the unabsorbed pion-exchange prediction and \OPEOverAbsorbed{} of the absorbed one.}
\label{tab:classes}
\end{table}

Three of the fitted values carry physics statements. The non-diffractive mass law is fitted at $\epsilon_{\rm nd} = \ParEpsNd$. The non-diffractive mass spectrum therefore falls more steeply than the diffractive one, whose exponent is the Pomeron value $\epsilon = 0.0808$. The one-pion exchange accounts for $s_\pi = \ParSPi$ of the single-diffraction class, \OPEMbPerSide{}~mb per side. The same flux predicts \OPEUnabsorbedMb{}~mb without absorption and \OPEAbsorbedMb{}~mb with the eikonal screening, shares of the single-diffraction class of \OPEUnabsorbedShare{} and \OPEAbsorbedShare{}. The fitted cross section lies between the two predictions. The absorbed prediction falls short of the fitted cross section by a factor of \OPEOverAbsorbed. The shape of the absorbed flux, which hardens the leading nucleons in $x_L$ and softens them in $p_T$, is what the spectra select. The flavor weights are $\gamma_s = \ParGammaS$ per created $s\bar s$ pair, $r_V = \ParRV$ for a vector against a pseudoscalar, $r_{B\bar B} = \ParRBB$ for a baryon pair at a continuing step, and the associated-production weight $a_{YK} = \ParAYK$ at the non-strange baryonic vertices. With the hyperon resonances and the excited kaons included in the catalog, $a_{YK}$ lies \AYKRelation.

\subsection{Spectra by species}\label{sec:spectra}

\begin{figure}[htbp]
\centering
\includegraphics[width=0.78\textwidth]{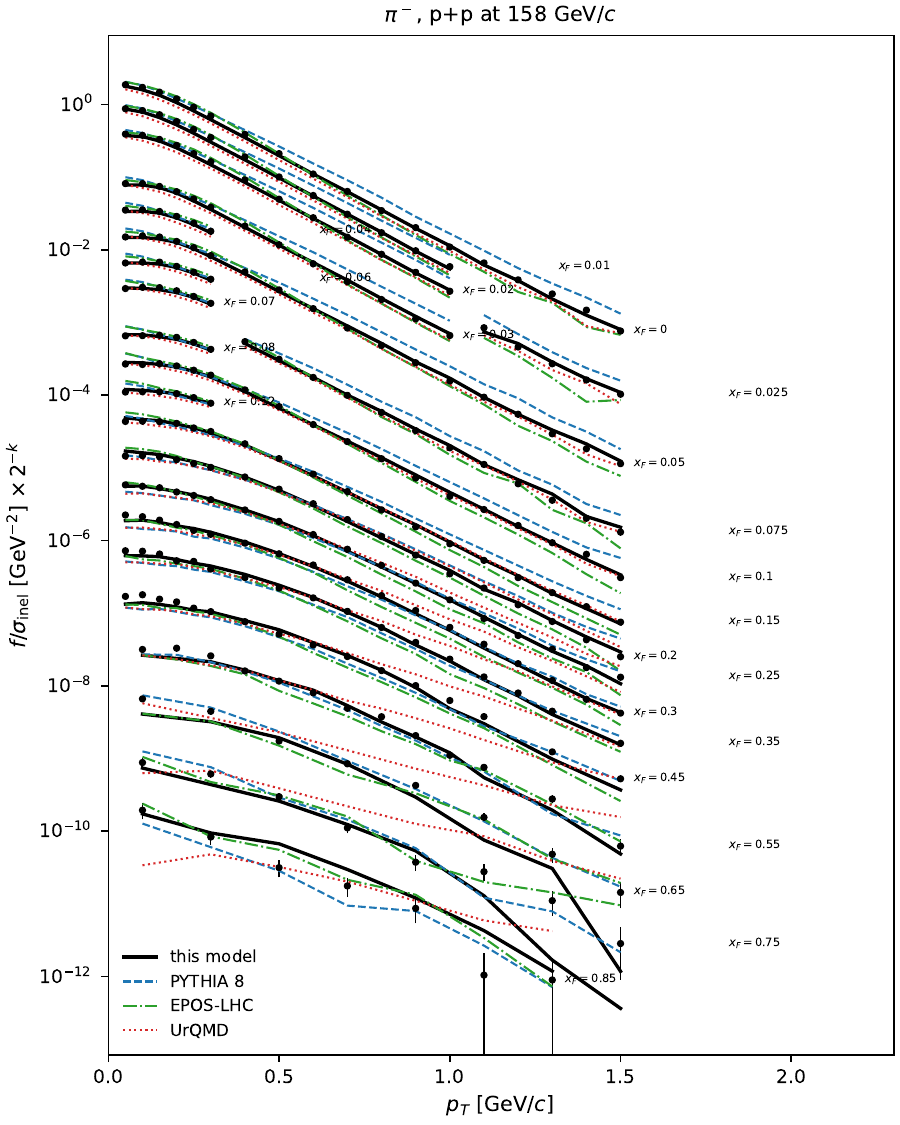}
\caption{Invariant cross section per inelastic event $f/\sigma_{\rm inel}$ of $\pi^-$ in p+p collisions at 158~GeV/$c$ against $p_T$ at fixed $x_F$. Points: NA49~\cite{na49pppi2006} with statistical errors. Full lines: this model at its fitted parameters, before the profiled scales of Table~\ref{tab:scales}. These scales are normalizations of \NormScaleRange\% for the six species and, for the kaons beyond $x_F = 0.25$, the identification scales, which lift the forward $K^+$ by \KpLiftPctRange\%. Broken lines: PYTHIA~8 (dashed), EPOS-LHC (dash-dotted), and UrQMD (dotted) at their published tunes, binned by the same estimator. Successive $x_F$ slices are multiplied by $1/2$ for clarity.}
\label{fig:piminus}
\end{figure}

\begin{figure}[htbp]
\centering
\includegraphics[width=0.78\textwidth]{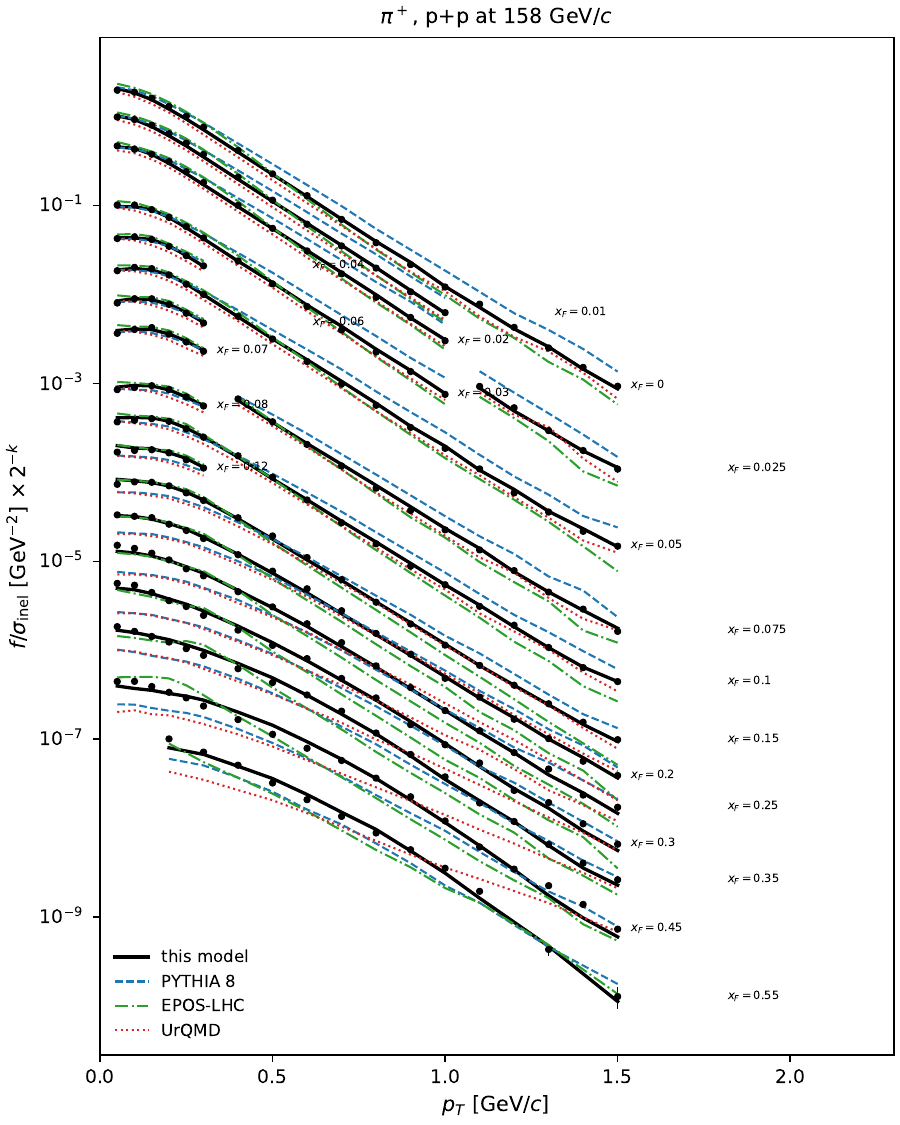}
\caption{As Figure~\ref{fig:piminus} for $\pi^+$.}
\label{fig:piplus}
\end{figure}

\begin{figure}[htbp]
\centering
\includegraphics[width=0.49\textwidth]{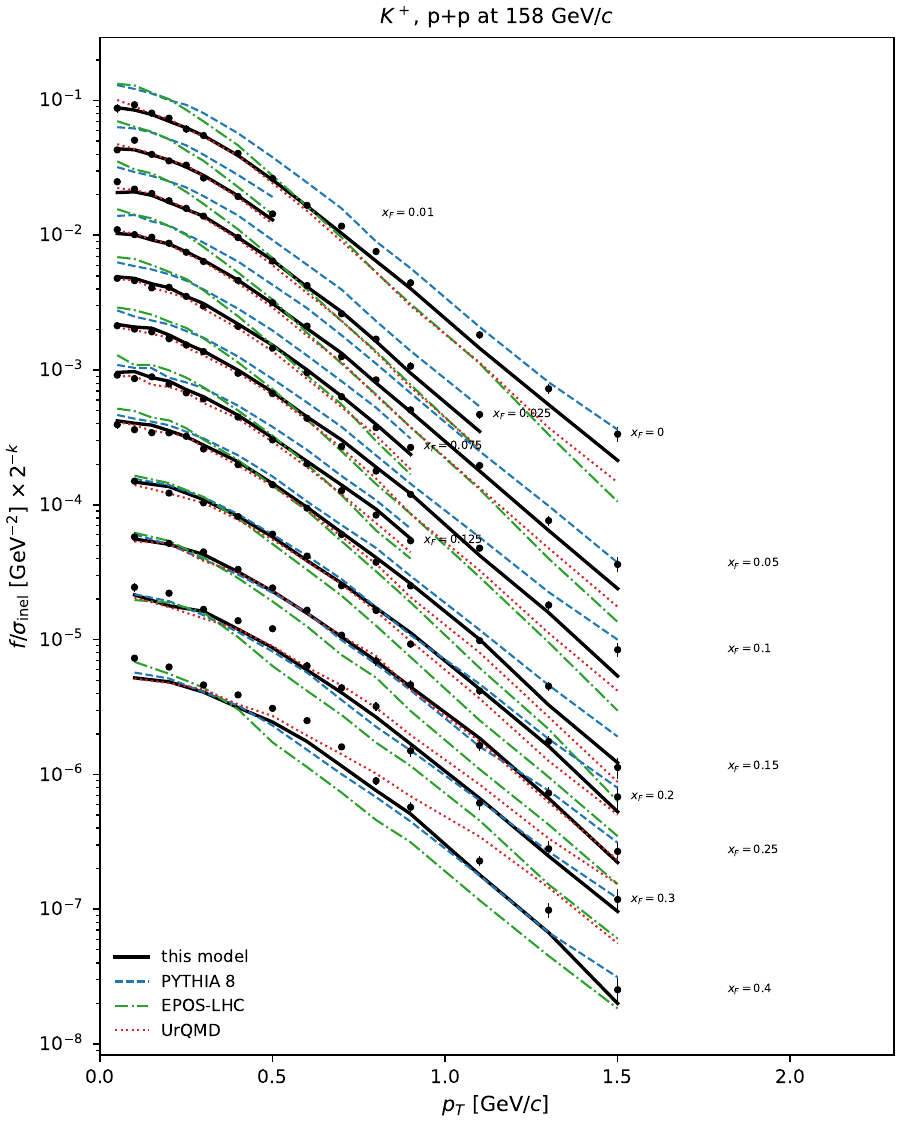}
\includegraphics[width=0.49\textwidth]{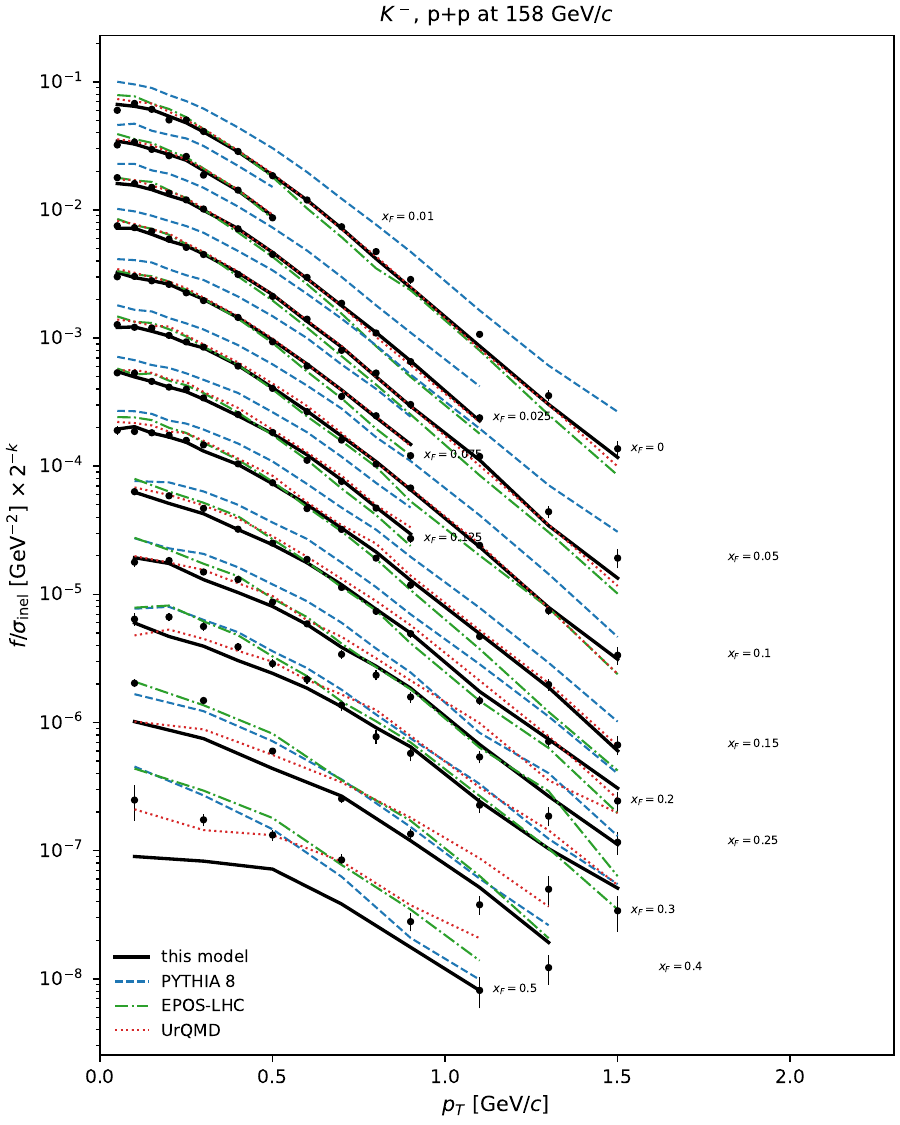}
\caption{As Figure~\ref{fig:piminus} for $K^+$ (left) and $K^-$ (right)~\cite{na49ppk2010}.}
\label{fig:kaons}
\end{figure}

\begin{figure}[htbp]
\centering
\includegraphics[width=0.78\textwidth]{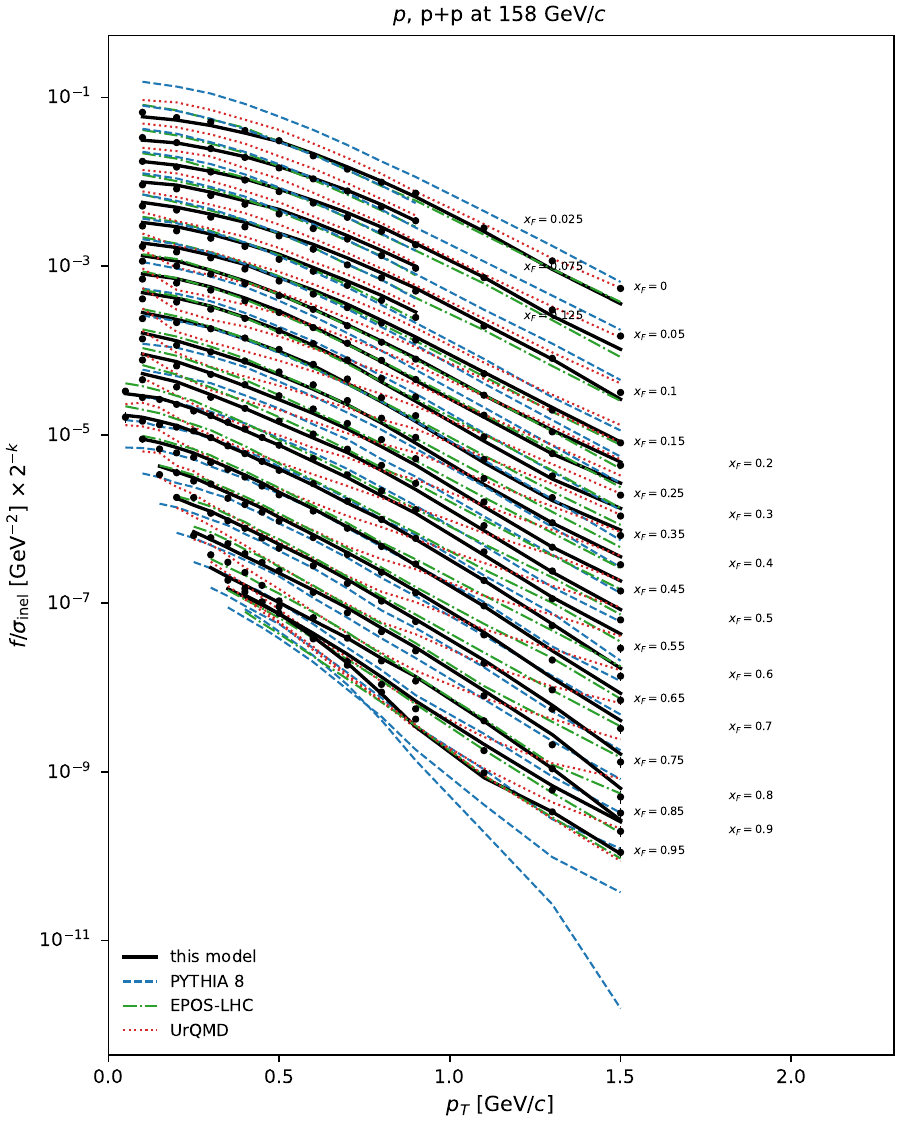}
\caption{As Figure~\ref{fig:piminus} for protons. The proton tables extend to $x_F = 0.95$, beyond the pion tables. The region $p_T < 0.4$~GeV/$c$ at $x_F > 0.6$ lies in the trigger veto and is not measured.}
\label{fig:proton}
\end{figure}

\begin{figure}[htbp]
\centering
\includegraphics[width=0.6\textwidth]{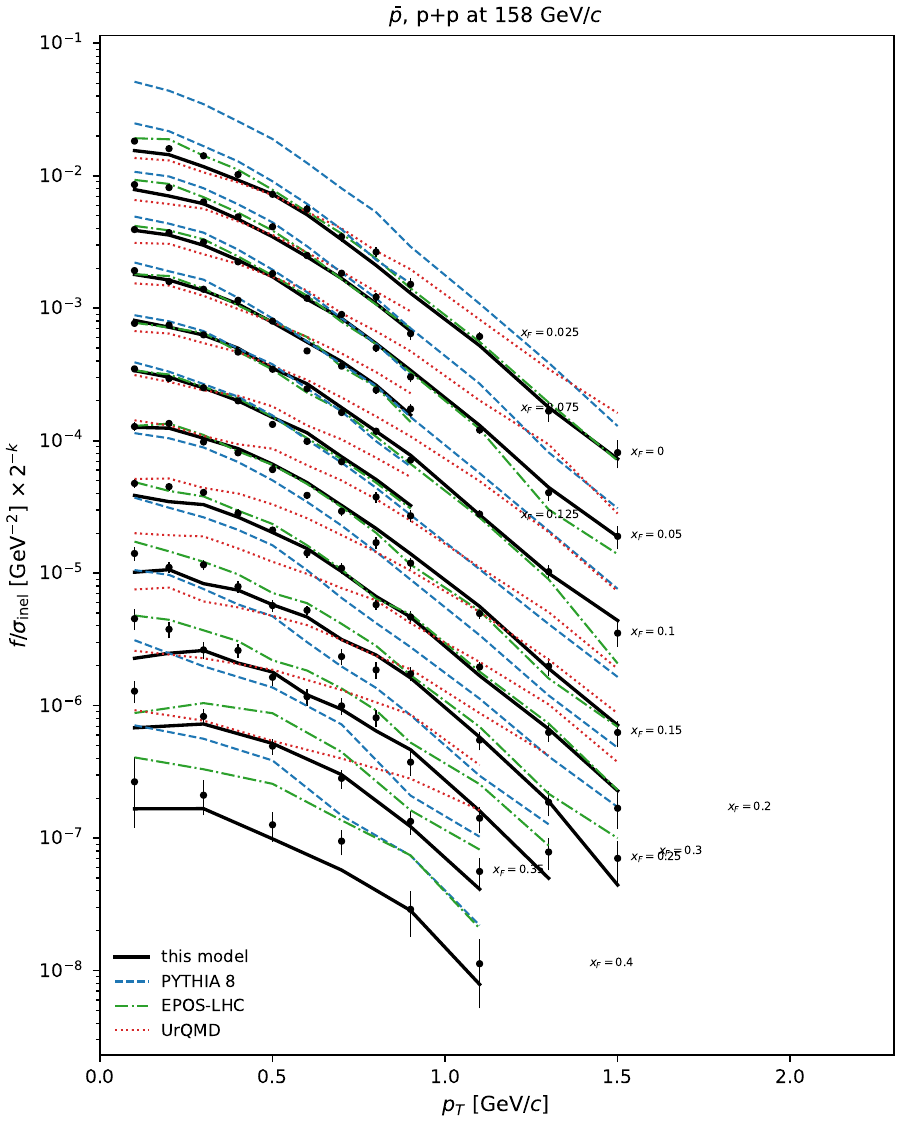}
\caption{As Figure~\ref{fig:piminus} for antiprotons.}
\label{fig:antiproton}
\end{figure}

\begin{figure}[htbp]
\centering
\includegraphics[width=\textwidth]{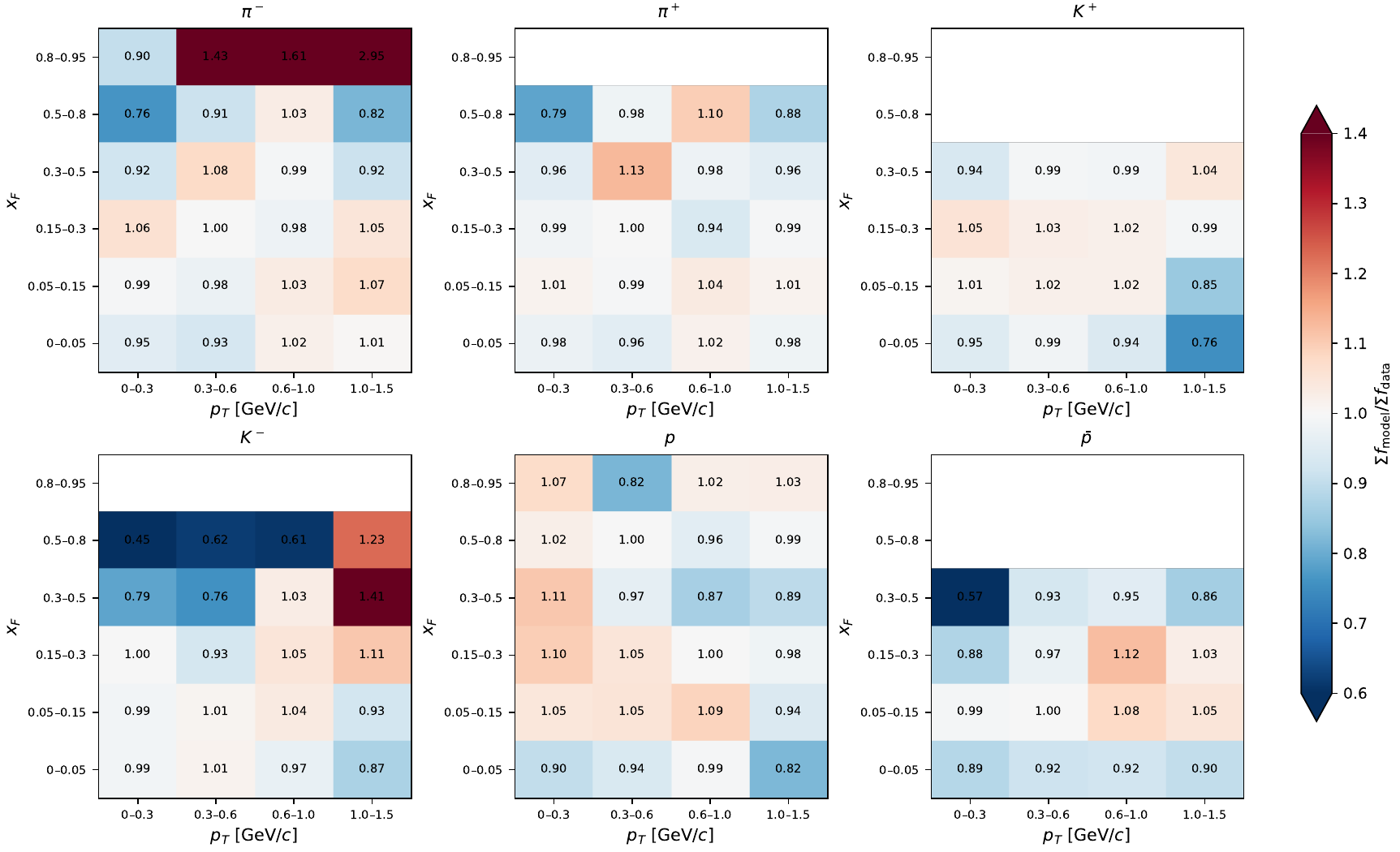}
\caption{The model under its profiled scales over the data in regions of the $(x_F, p_T)$ plane for the six species. Each cell is $\Sigma f_{\rm model} / \Sigma f_{\rm data}$ over the tabulated points inside it, so the largest cross sections of a cell set its value. The cell value is a ratio of summed cross sections, unweighted by $\chi^2$. The last column holds the points from 1.0 to 1.5~GeV/$c$, where every fitted table ends. The numbers in the cells are the ratios. The color scale stops at 0.6 and 1.4, and cells beyond it carry the end colors. The cells of the $\pi^-$ row at $x_F = 0.85$ hold one or two points each, with statistical errors of 22\% to 100\%.}
\label{fig:map}
\end{figure}

Figures~\ref{fig:piminus} to~\ref{fig:antiproton} show the six spectra in $x_F$ slices against $p_T$, with the model at its fitted parameters before the profiled scales. Figure~\ref{fig:map} shows the model under its profiled scales over the data in regions of the $(x_F, p_T)$ plane. Table~\ref{tab:segments} divides the $\chi^2$ by species.

\begin{table}[htbp]
\centering
\begin{tabular}{lcccc}
\toprule
Species & Points & $\chi^2$ & $\chi^2$ per point & Total yield, model / data \\
\midrule
\SegmentRows
\bottomrule
\end{tabular}
\caption{The $\chi^2$ by species at the minimum (the data term under the profiled scales, with the scale penalty of \NuisancePenalty{} added to the total), and the total yields per inelastic event of the model over the measured ones~\cite{spshadrons2026}.}
\label{tab:segments}
\end{table}

The protons account for the largest share of the $\chi^2$, most of it in the forward region. At $x_F > 0.5$ the hard protons above 1~GeV/$c$ stand at \RegPHardFwd{} of the data over \RegPHardFwdN{} points, with $\chi^2 = \RegPHardFwdChi$. Their $x_F$ shape at fixed $p_T$ is finer than the model's statistical resolution there. The soft protons at 0.3 to 0.6~GeV/$c$ stand at \RegPSoftMid{} at $x_F = 0.5$ to 0.8 and at \RegPSoftVfwd{} at $x_F = 0.8$ to 0.95 ($\chi^2 = \RegPSoftVfwdChi$ over \RegPSoftVfwdN{} points). The mid protons at $x_F = 0.30$ to $0.50$ and 0.6 to 1.0~GeV/$c$ stand at \RegPMid{} ($\chi^2 = \RegPMidChi$ over \RegPMidN{} points). Above 1~GeV/$c$ at $x_F < 0.15$ the protons stand at \RegPHardCentA{} and \RegPHardCentB{} of the data ($\chi^2 = \RegPHardCentChi$ over \RegPHardCentN{} points).

The pions account for the next share of the $\chi^2$. The soft $\pi^-$ at $x_F = 0.15$ to $0.30$ below 0.3~GeV/$c$ stand at \RegPimSoftMid{} of the data ($\chi^2 = \RegPimSoftMidChi$ over \RegPimSoftMidN{} points). The central soft $\pi^-$ at $x_F < 0.05$ stand at \RegPimSoftCent{} ($\chi^2 = \RegPimSoftCentChi$ over \RegPimSoftCentN{} points). The $\pi^+$ at $x_F = 0.30$ to $0.50$ and 0.3 to 0.6~GeV/$c$ stand at \RegPipMid{} ($\chi^2 = \RegPipMidChi$ over \RegPipMidN{} points). The very forward soft $\pi^-$ at $x_F = 0.5$ to 0.8 stand at \RegPimVfwdA{} below 0.3~GeV/$c$ and \RegPimVfwdB{} from 0.3 to 0.6~GeV/$c$ ($\chi^2 = \RegPimVfwdChi$ over \RegPimVfwdN{} points). The $\pi^+$ there stand at \RegPipVfwd{} below 0.3~GeV/$c$.

The kaons follow. Below 1~GeV/$c$ at $x_F \le 0.2$ the $K^+$ and $K^-$ stand at \RegKpCent{} and \RegKmCent{} of the data (\RegKpCentN{} points each). Above 1~GeV/$c$ at $x_F < 0.15$ the $K^+$ stand at \RegKpHardCentA{} and \RegKpHardCentB{} of the data ($\chi^2 = \RegKpHardCentChi$ over \RegKpHardCentN{} points). Beyond $x_F = 0.25$ the two charges differ:
\begin{enumerate}
\item The bare $K^+$ stand at \RegKpBareA{} of the data at $x_F = 0.30$ and at \RegKpBareB{} at $x_F = 0.40$. The identification scale of Table~\ref{tab:scales} lifts them to \KpLiftRange{} of the data in every $p_T$ column of Figure~\ref{fig:map}, at a pull of \KplusIdPull{} times the stated amplitude. The forward $K^+$ residual is therefore an offset in normalization, common to all $p_T$ at a given $x_F$.
\item The forward $K^-$ show a residual in the shape of their $p_T$ spectrum, which a scale that depends on $x_F$ alone cannot remove. After their scale, at $x_F = 0.30$ to $0.50$, they stand at \RegKmLowA{} and \RegKmLowB{} of the data below 0.6~GeV/$c$ and at \RegKmHigh{} above 1~GeV/$c$. In the $x_F = 0.5$ row, whose cells hold one or two points, they stand at \RegKmRowLowRange{} and \RegKmRowHigh{}. The \RegKmFwdN{} points at $x_F \ge 0.30$ contribute $\chi^2 = \RegKmFwdChi$ of the $K^-$ total of \KmSegChi.
\end{enumerate}
The antiprotons contribute $\chi^2 = \ChiPerPointAntiproton$ per point, below the fit's average of \FitSsq.

\subsection{Systematic scales and variants}\label{sec:scales}

Table~\ref{tab:scales} lists the profiled scales at the minimum. Of the six normalizations, \NormWithinOne{} stay within 1.0 of their 1.5\% priors. The proton and antiproton normalizations have pulls of $\NormPullProton$ and $\NormPullAntiproton$. The neutron scale is fitted at $\NeutronScalePct\%$, a pull of \NeutronScalePull{} times its prior. The model's leading neutrons therefore exceed the data by a common shift of \NeutronScalePull{} times the calorimeter-scale prior (Section~\ref{sec:neutrons}). The $K^+$ identification scale is fitted at \KplusIdPull{} times the stated amplitude. The model's $K^+$ above $x_F = 0.25$ fall below the data by \KplusIdPull{} times the identification uncertainty the kaon paper states. That is the one normalization residual beyond the data's own systematics (Section~\ref{sec:strangeness}). The $K^-$ identification scale is fitted at \KminusIdPull{} times the stated amplitude, a lift of at most \KminusIdLift\%. The forward $K^-$ residual lies in the shape of the $p_T$ spectrum, which a scale common to all $p_T$ at one $x_F$ cannot absorb (Section~\ref{sec:spectra}).

\begin{table}[htbp]
\centering
\begin{tabular}{lcc}
\toprule
Scale (prior) & Value & Pull \\
\midrule
\NuisanceRows
\bottomrule
\end{tabular}
\caption{The profiled systematic scales at the minimum. A positive value scales the model up to meet the data. The kaon identification scales are given in units of their $x_F$-dependent amplitude.}
\label{tab:scales}
\end{table}

The systematic variants of the comparison move the $\chi^2$ as Table~\ref{tab:variants} lists. Under the normalization variants every parameter stays within \VarNormMaxMove{} errors. The diffractive cross-section variants move \VarXsecMovedPars{} by up to \VarXsecMaxMove{} errors, since these parameters describe components of the single-diffraction class, whose total size the anchor sets. Changing the discrepancy floor rescales the $\chi^2$ and moves the parameters by at most \VarFloorMaxMove{} errors.

\begin{table}[htbp]
\centering
\begin{tabular}{lc}
\toprule
Variant & $\Delta\chi^2$ \\
\midrule
\VariantRows
\bottomrule
\end{tabular}
\caption{The systematic variants of the comparison, refitted, against the nominal minimum. The discrepancy floor takes 3\%, 7\%, and 10\% in place of 5\%. The common normalization shift multiplies every prediction of the model by $1 \pm 0.02$ and leaves the priors of the species scales at 1.5\%. The two diffractive variants move the single-diffraction anchor by $\pm 0.8$~mb and the double-diffraction cross section to 0.5 and 2.5~mb.}
\label{tab:variants}
\end{table}

\subsection{Leading neutrons}\label{sec:neutrons}

\begin{figure}[htbp]
\centering
\includegraphics[width=0.5\textwidth]{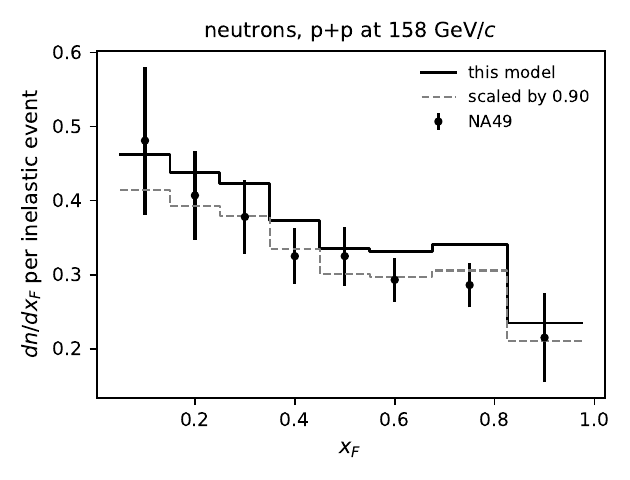}
\caption{The neutron $dn/dx_F$ per inelastic event: NA49 (points, systematic errors) and the model before (full) and under (broken) its profiled scale of $\NeutronScalePct\%$.}
\label{fig:neutrons}
\end{figure}

The neutron $dn/dx_F$ constrains two sources together, the one-pion exchange and the decays of excited proton lines (Table~\ref{tab:neutrons}, Figure~\ref{fig:neutrons}). The model's neutrons stand at \NeutronRatioLow{} of the data at $x_F = 0.1$ and rise to \NeutronRatioMid{} at 0.4 and \NeutronRatioMax{} at \NeutronRatioMaxXf. With the common scale of $\NeutronScalePct\%$ the eight points yield $\chi^2 = \NeutronChi$. The leading neutrons of the cascade come from the $n\pi^+$ decays of the low-mass $\Delta^+$ and $N^{*+}$ lines, whose Clebsch--Gordan shares leave no freedom. They also come from the exchange at its fitted share. At $x_F = 0.6$ to $0.9$ the comparison is limited by the data's correlated systematic uncertainties, which exceed the statistical ones there. The neutron paper states that the 10\% to 28\% per point represent the calorimeter's energy scale and unfolding~\cite{na49ppproton2010}. The ISR neutron spectra at nonzero production angles report invariant cross sections that scale in $x_F$, with a large-$x_F$ shape that carries the pion-exchange contribution~\cite{engler1975neutron}. The zero-degree ISR spectra report the same scaling and large-$x_F$ behavior at $p_T = 0$~\cite{flauger1976neutron}. Double-differential neutron cross sections in p+p also exist at 12 and 24~GeV/$c$~\cite{blobel1978neutron}. Fermilab measured inclusive neutrons from 400~GeV protons on hydrogen at laboratory angles 0.7 to 10~mrad~\cite{whalley1979neutron}. The NA49 proton paper treats that Fermilab set as the double-differential measurement in the SPS energy range. It extracts $p_T$-integrated yields from that set and from the ISR yields, and it concludes that those yields are not a basis for a quantitative comparison~\cite{na49ppproton2010}. The forward-neutron $dn/dx_F$ of the Main Injector Particle Production experiment (MIPP) at 58 and 84~GeV/$c$ agrees with the NA49 table at $x_F > 0.6$. At lower $x_F$ it lies about a factor of two above that table, a discrepancy MIPP leaves unexplained~\cite{mipp2011neutron}. The MIPP calorimeter had no transverse segmentation, so every MIPP cross section is integrated over $p_T$. The $p_T$ shape of the model's neutrons awaits a fixed-target measurement at SPS energies with a transversely segmented calorimeter.

\begin{table}[htbp]
\centering
\begin{tabular}{lccc}
\toprule
$x_F$ & NA49 $dn/dx_F$ & Model & Model / data \\
\midrule
\NeutronRows
\bottomrule
\end{tabular}
\caption{The neutron $x_F$ distribution per inelastic event in the forward hemisphere, data and the model before its profiled scale. The data's errors are the systematic uncertainties of the calorimeter measurement.}
\label{tab:neutrons}
\end{table}

\subsection{Transverse momentum and multiplicity}\label{sec:seagull}

\begin{figure}[htbp]
\centering
\includegraphics[width=\textwidth]{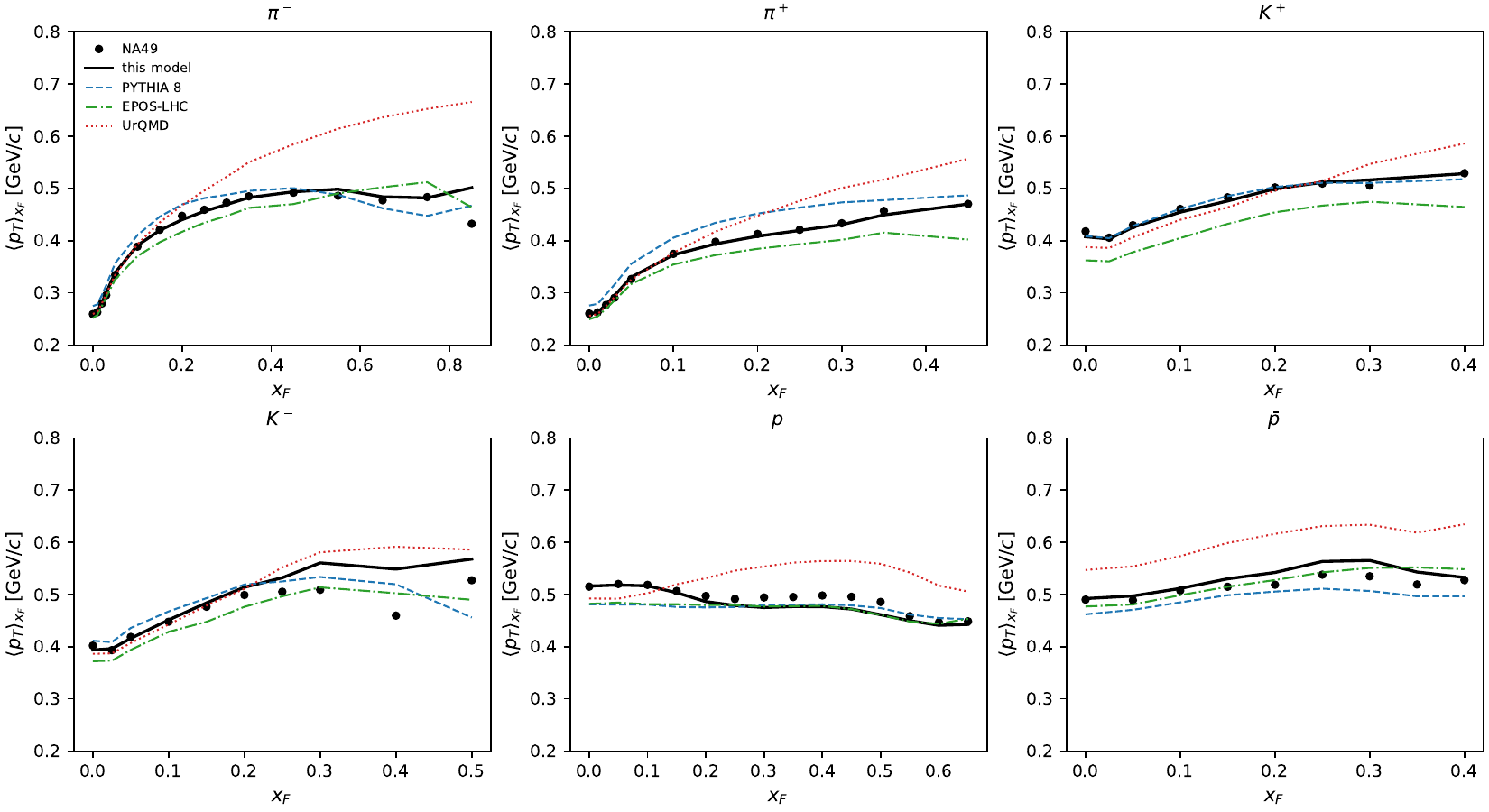}
\caption{The seagull effect: $\langle p_T\rangle_{x_F}$ against $x_F$ for the six species in p+p collisions at 158~GeV/$c$, from the tabulated invariant cross sections with the quadrature of Section~\ref{sec:seagull} on the $x_F$ rows with full transverse support. Points: NA49~\cite{na49pppi2006,na49ppk2010,na49ppproton2010}. Full lines: this model. Broken lines: PYTHIA~8 (dashed), EPOS-LHC (dash-dotted), and UrQMD (dotted), binned by the same estimator. The normalizations cancel in the ratio, so $f$ and $f/\sigma_{\rm inel}$ give the same profile.}
\label{fig:seagull}
\end{figure}

Two readouts of the fitted model test the decay rule. The first is the seagull effect, the rise of the mean transverse momentum with $x_F$~\cite{bardadin1966seagull,vanapeldoorn1975seagull}. It reuses the tabulated spectra of the fit, so it is a derived diagnostic of the fitted spectra. It is evaluated for the data and for every model with the same quadrature over the same tabulated points,
\begin{equation}
\langle p_T\rangle_{x_F} = \frac{\sum_i p_{T,i}\,(p_{T,i}/E_i)\,f_i\,\Delta p_{T,i}}{\sum_i (p_{T,i}/E_i)\,f_i\,\Delta p_{T,i}},
\end{equation}
with $f_i$ the invariant cross section, $E_i$ the c.m.\ energy at the point, and $\Delta p_{T,i}$ the midpoint bin widths. Only the $x_F$ rows whose tabulated $p_T$ range runs from 0.1 to at least 0.95~GeV/$c$ enter, since a truncated row would shift the quadrature for the data and the models alike. The per-species normalizations and the identification scale of the kaons depend on the species and on $x_F$ only, so they cancel in the ratio. The profiled scales of the comparison therefore do not enter. Figure~\ref{fig:seagull} shows the profiles, and Table~\ref{tab:seagull} the intercepts and the deviations. Several features are noteworthy:
\begin{enumerate}
\item The model's intercepts at $x_F = 0$ run from \SeagullInterceptPion~GeV/$c$ for the pions to \SeagullInterceptProton{} for the protons. They follow the data's to within \SeagullInterceptMaxDev~GeV/$c$. Heavier daughters of sequential two-body decays carry harder transverse spectra, so the hierarchy follows from the decay chain without a parameter of its own.
\item The rise with $x_F$ is reproduced for $\pi^+$ and $K^+$ to rms deviations of \SeagullRmsPiPlus{} and \SeagullRmsKPlus~GeV/$c$, and for $\pi^-$ and $\bar p$ to \SeagullRmsPiMinus{} and \SeagullRmsAntiproton. The very forward $\pi^-$ at $x_F = \SeagullPimFwdXf$ lie \SeagullPimFwd~GeV/$c$ above the data. The protons follow the measured plateau and its fall beyond $x_F = 0.5$, up to \SeagullProtonBelow~GeV/$c$ below the data.
\item The $K^-$ beyond $x_F = 0.3$ lie \SeagullKminusFwd~GeV/$c$ above the data. This offset is the transverse shape of the forward $K^-$ (Section~\ref{sec:spectra}), below the data under 0.6~GeV/$c$ and above it beyond 1~GeV/$c$.
\item Of the generators, PYTHIA~8 reproduces the $K^+$ profile and EPOS-LHC the $K^-$ profile most closely. EPOS-LHC lies 0.05~GeV/$c$ below the data on the $K^+$. UrQMD lies 0.03 to 0.09~GeV/$c$ above the data on the pions and on the baryons.
\end{enumerate}
The transverse scales the fit sets are the vertex slope $b$ and the intact proton's vertex term $f$. The species hierarchy and the slopes of the profiles are consequences of the decay chain.

\begin{table}[htbp]
\centering
\small
\setlength{\tabcolsep}{4pt}
\begin{tabular}{lrcc cccc cccc}
\toprule
 & & \multicolumn{2}{c}{$x_F = 0$} & \multicolumn{4}{c}{rms deviation} & \multicolumn{4}{c}{mean deviation} \\
\cmidrule(lr){3-4}\cmidrule(lr){5-8}\cmidrule(lr){9-12}
 & rows & data & model & model & PY8 & EPOS & UrQMD & model & PY8 & EPOS & UrQMD \\
\midrule
\SeagullRows
\bottomrule
\end{tabular}
\caption{The seagull of Figure~\ref{fig:seagull} in numbers: the number of $x_F$ rows with full transverse support, $\langle p_T\rangle_{x_F}$ at $x_F = 0$ for the data and this model, and the rms and mean deviation of each model's $\langle p_T\rangle_{x_F}$ from the data's over the rows, all in GeV/$c$. The quadrature is the same for the data and every model.}
\label{tab:seagull}
\end{table}

\begin{figure}[htbp]
\centering
\includegraphics[width=\textwidth]{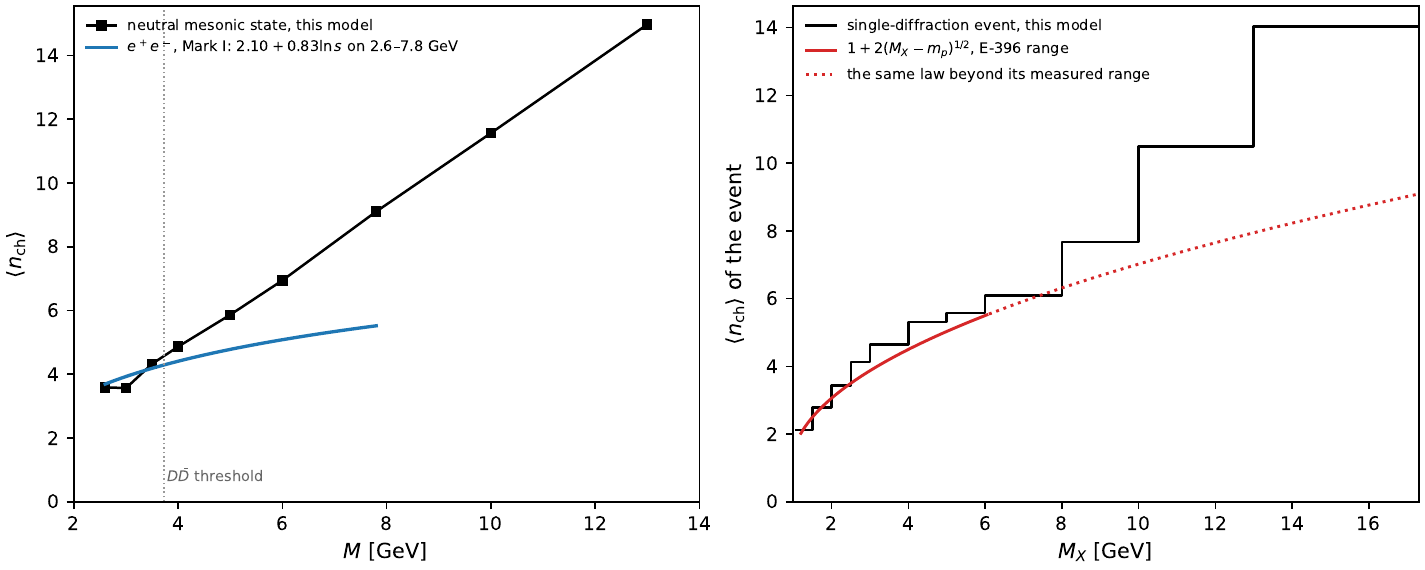}
\caption{Left: the mean charged multiplicity of a neutral mesonic continuum state of mass $M$, decayed at rest through the cascade at the fitted point with $10^3$ decays per mass (squares), and the $e^+e^-$ annihilation multiplicity $\langle n_{\rm ch}\rangle = 2.10 + 0.83\ln s$, a fit of this paper to the Mark~I data, drawn over the measured range of 2.6 to 7.8~GeV (full line)~\cite{siegrist1982markI}. The dotted vertical line marks the open-charm threshold. Right: the whole-event charged multiplicity of the model's single-diffraction events against the excited mass $M_X$ (histogram), and the E-396 law $2\,(M_X - m_p)^{1/2}$ plus one charged particle for the leading proton, full over its measured range and dotted beyond~\cite{goulianos1982universality}. The charged products of the $K^0_S$ and $\Lambda$ decays are counted in both panels.}
\label{fig:nchmass}
\end{figure}

The second readout is the charged multiplicity of an excited system against its mass, the observable that separates a cascade from a string at high mass. A neutral mesonic continuum state of mass $M$ is decayed at rest through the cascade with the fitted flavor weights, with $10^3$ decays per mass. The charged products of the $K^0_S$ and $\Lambda$ decays are counted as the measurement counts them. Its reference is the mean charged multiplicity of $e^+e^-$ annihilation into hadrons, measured by Mark~I at 38 energies between 2.6 and 7.8~GeV~\cite{siegrist1982markI}. A straight line in $\ln s$ fitted here to those points gives $\langle n_{\rm ch}\rangle = (2.10 \pm 0.12) + (0.832 \pm 0.037)\ln s$ with $s$ in GeV$^2$, at $\chi^2/{\rm ndf} = 30.5/36$. Two measurements support the use of a quark--antiquark system as the reference for a color-singlet hadronic state. The $gg$ system of $\Upsilon(1S) \to gg\gamma$ carries the charged multiplicity of the $q\bar q$ system of the same mass to a ratio of $1.04 \pm 0.02 \pm 0.05$ below 7~GeV~\cite{cleo1997gluonquark}. The diffractive and non-diffractive systems of hadron collisions follow one universal function of the mass available for pionization~\cite{goulianos1982universality}. Figure~\ref{fig:nchmass} shows the result. Three facts stand out.
\begin{enumerate}
\item Below the open-charm threshold at 3.73~GeV the cascade's count stands at \NchLowRange{} of the Mark~I line. At 2.6~GeV the state makes \NchAtTwoSix{} charged hadrons against \NchAtTwoSixRef.
\item Above it the count grows linearly, by \NchSlope{} charged hadrons per GeV. The measured multiplicity grows with $\ln s$. At 5~GeV the cascade gives \NchAtFive{} against \NchAtFiveRef, and at 7.8~GeV, the top of the measured range, \NchAtSevenEight{} against \NchAtSevenEightRef, a factor of \NchAtSevenEightRatio. Above the threshold the Mark~I multiplicity includes the decays of charmed hadrons, which the cascade does not make. The comparison there sets the linear growth of the continuation step against a logarithmic one. It does not compare the cascade with light-flavor data.
\item The single-diffraction systems are the cleaner comparison (right panel), since the measured systems are themselves hadronic excitations of the proton. Fermilab E-396 measured $\langle n\rangle \approx 2\,(M_X - m_p)^{1/2}$ for the diffractive states of $h\,p \to X\,p$ up to $M_X \approx 6$~GeV~\cite{goulianos1982universality}. The charged multiplicity of the model's excited system, that is, of the whole event less the leading proton, is \SdLawInside{} times that law at $M_X = 2.5$ to 6~GeV. Beyond that range it is \SdLawBeyondText.
\end{enumerate}
The linear growth is a property of the continuation step. The density $\exp(m/T_H)$ keeps each continuation daughter within a few $T_H$ of the open edge. A heavy state therefore sheds its mass in many steps of a fixed momentum scale, as a Hagedorn fireball does. The same soft steps populate the central region of the spectra. A continuation scale that grows with the mass brings the count toward the $e^+e^-$ line. It also hardens the central hadrons of every species, at a cost of \HeavyMassScaleCost{} in $\chi^2$ (Section~\ref{sec:strangeness}). The spectra integrate over the mass distribution of the excited systems. With the fitted mass laws the model gives \NchEventModel{} charged hadrons per inelastic event, against the measured \NchEventData{} summed over the six species. The total yields cannot separate the multiplicity--mass law from the mass spectrum, since a change in one can be compensated by the other. The multiplicity of the heavy systems is therefore where the cascade departs from a measured law. An event-level measurement of the excited mass would decide between the cascade's count and the measured one (Section~\ref{sec:discussion}).

\subsection{Generators on the same points}\label{sec:genresult}

Table~\ref{tab:generators} gives the $\chi^2$ of the three fixed-tune generators and of this model on the same 1,254 points under the same error model. It also gives the yields in the measured window. Each species column is the data term at that model's own profiled scales, which Table~\ref{tab:genpulls} lists. A negative normalization pull scales the model down. A species column after a large pull therefore measures the shape left once the yield has been rescaled. The penalty for the rescaling enters only the total. Several features are noteworthy.
\begin{enumerate}
\item EPOS-LHC comes closest of the generators, at \GenRatioEpos{} times this model's $\chi^2$, with the window yields within 7\% of the data. Its largest $\chi^2$ per point lies in the $\pi^+$ and the $K^+$, 10.7 and 10.6. Its $K^+$ normalization and identification scales there are $-8.0$ and $+4.9$ priors. Its protons follow at 8.1 per point.
\item PYTHIA~8 and UrQMD stand at \GenRatioPythia{} and \GenRatioUrqmd{} times this model's $\chi^2$. PYTHIA~8's protons contribute $\chi^2 = 20{,}897$ at a window yield of 0.94, a failure of shape at a nearly correct integral. Its antiprotons, at 2.46 of the data in the window, are scaled down by 55\% (a pull of $-37$). That pull alone adds about 1,360 to its total. UrQMD's protons and antiprotons are shape failures, with $\chi^2 = 13{,}983$ and 2,217, the latter at a window yield of 1.20. Its pions stand at 0.86 and 0.84 of the data in the window.
\item The ranking does not depend on the profiled scales. Without the penalties the species sums are \GeneratorSpeciesSums{} for PYTHIA~8, EPOS-LHC, and UrQMD, against \CascadeSpeciesSum{} for this model.
\item This model's $\chi^2$ per point is \ChiPerPointKplus{} for the $K^+$, \ChiPerPointAntiproton{} for the antiprotons, \ChiPerPointKminus{} for the $K^-$, and \ChiPerPointPionProtonRange{} for the pions and protons, whose statistical errors are smallest. The $K^-$ value is the forward transverse shape of Section~\ref{sec:spectra}.
\end{enumerate}
The generators describe the central pions to within 10\%. They differ from the data most where the hadronization of the excited remnant decides, namely in the forward protons for PYTHIA~8 and UrQMD, in the antiprotons for UrQMD, and in the $K^+$ for EPOS-LHC. The ranking describes the published tunes at this energy. What the generators could reach with retuning lies outside it.

\begin{table}[htbp]
\centering
\small
\begin{tabular}{lccccccc}
\toprule
Model & $\pi^-$ & $\pi^+$ & $K^+$ & $K^-$ & $p$ & $\bar p$ & Total \\
\midrule
\GeneratorRows
\midrule
\multicolumn{8}{l}{Yield in the measured window, model / data} \\
\GeneratorSumRows
\bottomrule
\end{tabular}
\caption{The $\chi^2$ of this model and of the fixed-tune generators on the 1,254 points of the six spectra under the error model of Section~\ref{sec:errors}, and the yield in the measured window of each model over the data. The normalization and kaon identification scales are profiled for each model (Table~\ref{tab:genpulls}). The species columns are the data terms at those scales, and the total adds their penalty. The generators ran $\GeneratorEvents \times 10^6$ inelastic events each. This model's row is the fit on the same points, without the neutron segment and its scale.}
\label{tab:generators}
\end{table}

\begin{table}[htbp]
\centering
\small
\setlength{\tabcolsep}{4pt}
\begin{tabular}{lccccccccc}
\toprule
 & \multicolumn{6}{c}{normalization pull} & \multicolumn{2}{c}{identification pull} & \\
\cmidrule(lr){2-7}\cmidrule(lr){8-9}
Model & $\pi^-$ & $\pi^+$ & $K^+$ & $K^-$ & $p$ & $\bar p$ & $K^+$ & $K^-$ & Penalty \\
\midrule
\GeneratorPullRows
\bottomrule
\end{tabular}
\caption{The profiled scales of each model on the points of Table~\ref{tab:generators}. The normalization pulls are in units of the 1.5\% prior, and the kaon identification pulls in units of the amplitude the kaon paper states. A negative pull scales the model down. The penalty is the sum of the squared pulls, which the totals of Table~\ref{tab:generators} include.}
\label{tab:genpulls}
\end{table}

\subsection{Cross-checks outside the fit}\label{sec:crosschecks}

The fit uses the reference-quality NA49 spectra only. The resonance and hyperon yields of other experiments, NA27 at 400~GeV/$c$, NA61/SHINE at 158~GeV/$c$, and NA49's own $K^*(892)$ and $\phi$, enter as cross-checks. Fischer et~al.\ find that the spectrometer experiments outside the reference set need normalization factors from 0.75 to 1.30~\cite{fischer2022piminus}. A yield is therefore compared here with an adopted allowance of 20\% to 30\% beside its quoted uncertainty. That allowance is transferred from the pion compilation and is not established for each measurement. A ratio measured within one experiment is the stronger comparison.

\paragraph{Resonance yields.} At the fitted point the model gives per inelastic event $\rho^0 = \RhoZero$ and $\omega = \OmegaYield$, against NA27's $0.385 \pm 0.018$ and $0.390 \pm 0.024$ at 400~GeV/$c$~\cite{na27resonances1991}. A ratio cancels the normalizations, so a ratio measured within one experiment is the sharper test. The model's $\rho^0/\omega = \RhoOverOmega$ compares with NA27's $0.99 \pm 0.08$, which NA60 confirmed with dimuons in p+A at 400~GeV/$c$ at $1.00 \pm 0.06$~\cite{na60rhoom2016}. This low ratio is a tension of the model with two experiments. It has a definite origin in the fit. The NA49 pion spectra hold the $\omega$ only through its charged pair. Indeed, they prefer the $\omega$'s three soft pions to the $\rho$'s two harder ones at every mesonic vertex. The model nevertheless keeps the common counting of the vector nonet in place of a factor on one state.

\begin{figure}[htbp]
\centering
\includegraphics[width=0.48\textwidth]{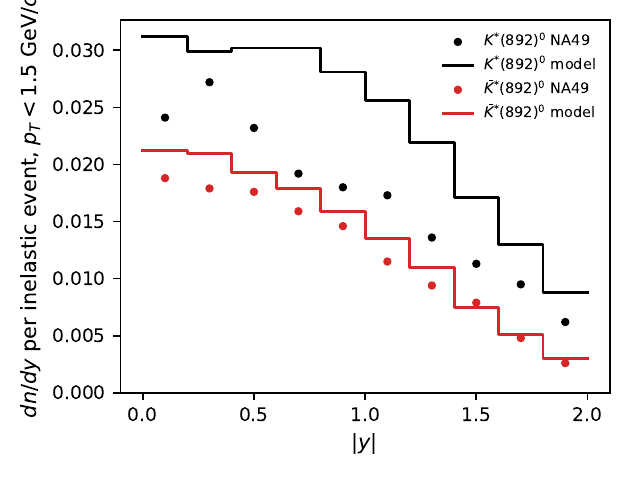}
\caption{The $K^*(892)^0$ and $\bar K^*(892)^0$ $dn/dy$ per inelastic event for $p_T < 1.5$~GeV/$c$, NA49 (points) and the model (histograms).}
\label{fig:kstar}
\end{figure}

The strange vectors show an excess as well. The model gives $\phi = \PhiYield$, against NA49's $0.0120 \pm 0.0015$~\cite{na49phi2000} and NA61/SHINE's $0.0126 \pm 0.0005$~\cite{na61phi2020}, which is \PhiRatioRange{} of these values. It lies \PhiSigmaNAfortynine{} standard deviations from NA49's own measurement, which shares the detector and normalization with the fitted spectra. The model gives $K^*(892)^0 = \KstarZero$ and $\bar K^*(892)^0 = \KstarZeroBar$, against NA49's $0.0741 \pm 0.0069$ and $0.0523 \pm 0.0048$~\cite{na49kstar2011}. Its ratio $K^{*0}/\bar K^{*0} = \KstarRatio$ compares with NA49's $1.42 \pm 0.04$. In NA49's rapidity bins the model's $K^{*0}$ stands at \KstarBinRange{} of the data, and at \KstarFwdBinRange{} in the forward bins from $|y| = 0.7$. The model's $\bar K^{*0}$ stands at \KstarBarBinRange{} (Figure~\ref{fig:kstar}). The $\phi$ is the hidden-strangeness member of the same vector excess as the forward $K^{*0}$. At the fitted $r_V$ and $\gamma_s$ a created $s\bar s$ pair leaves the vertex as a vector too often. The six spectra contain no $\phi$ and therefore do not constrain this. The two excesses are thus one statement about the vector weight on the strange channels. The $\phi$ yield is the measurement that would fix it. The excess of $K^{*0}$ over $\bar K^{*0}$ belongs, in addition, to the forward-strangeness residual of Section~\ref{sec:strangeness}.

\begin{figure}[htbp]
\centering
\includegraphics[width=0.48\textwidth]{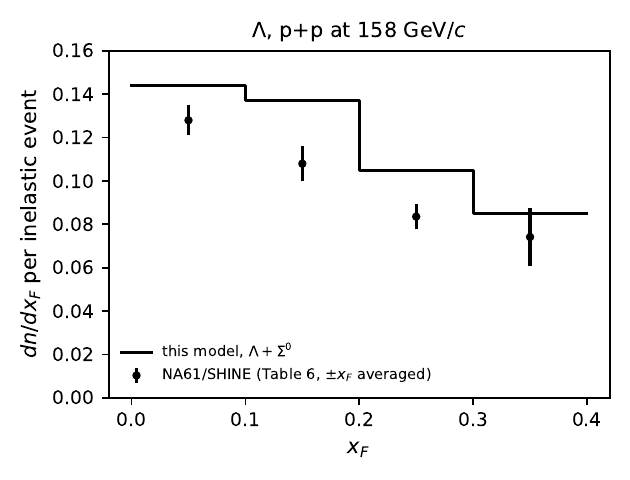}
\caption{The $\Lambda$ (with $\Sigma^0$) $dn/dx_F$ per inelastic event, NA61/SHINE~\cite{na61lambda2016} (points, the $\pm x_F$ rows averaged, statistical and systematic errors) and the model (histogram).}
\label{fig:lambda}
\end{figure}

\paragraph{$\Lambda$ distribution in $x_F$.} NA61/SHINE measured $dn/dx_F$ of the $\Lambda$ (including $\Sigma^0 \to \Lambda\gamma$) in p+p at 158~GeV/$c$ over $|x_F| < 0.4$~\cite{na61lambda2016}. It falls from 0.128 per event at $x_F = 0.05$ to 0.074 at 0.35. The model's $\Lambda + \Sigma^0$ is flat over that range, at \LambdaRatios{} of the data at $x_F = 0.05$, 0.15, 0.25, and 0.35. The total over the four windows is \LambdaTotal{} of the data (Figure~\ref{fig:lambda}). The forward excess comes from the leading-baryon split, which aligns the hyperon along the incoming proton. Without it the $\Lambda$ at $x_F = 0.3$ to $0.4$ follow the data, while the agreement of the NA49 $K^+$ and proton spectra worsens (Section~\ref{sec:strangeness}). The fit therefore keeps the alignment, which the NA49 spectra favor over the NA61/SHINE $\Lambda$ shape.

\subsection{What the spectra fix and what they leave open}\label{sec:strangeness}

Each component of Section~\ref{sec:model} was adopted against the alternatives that the data could distinguish from it. The alternatives are recorded with their cost. The following ones bound what a cascade of this kind can and cannot do.

\emph{The hadronization of the heavy systems.} The alternatives to the one-line cascade tested here for an excited system above a few GeV were evaluated on the same events. An isotropic split of the system into a baryonic and a mesonic cluster costs \HeavyClusterCost{} in $\chi^2$. A string-like step that emits one hadron at a time toward the center, while carrying the remaining mass forward, costs \HeavyStringCost{}. A two-line collision topology costs \HeavyTwoLineCost{}. A continuation step whose momentum scale grows with the mass brings the count of a heavy system toward the $e^+e^-$ value and costs \HeavyMassScaleCost{}. Among the forms tested, the spectra select the one-line cascade rule. A continuum state continues while a continuation channel is open and ends in one direct pair at the floor. A state of 8~GeV takes four to five steps, and four in five of its emitted siblings are resonances. The price of this rule is the multiplicity of the heavy systems, which grows linearly with the mass above 4~GeV and exceeds the $e^+e^-$ measurement there (Section~\ref{sec:seagull}). Decays of a floor state into three or more bodies add hard pions where the measured pions are soft. The data limit them to a few percent of the floor decays.

\emph{The soft pions.} The central soft $\pi^-$ stand at \RegPimSoftCent{} of the data and the mid soft $\pi^-$ at \RegPimSoftMid. Fischer et~al.\ isolate a central low-$p_T$ $\pi^-$ component with an inverse $m_T$ slope of about 0.08~GeV/$c$, independent of energy. They interpret it as further resonant states of low $Q$ value and small width~\cite{fischer2022piminus}. In the cascade the longitudinal position of the soft pions is set by the decay kinematics of the lines. The alignment of the continuing baryonic steps (Section~\ref{sec:cascade}) sends the hadrons a line emits toward the center. It places the soft $\pi^-$ where they stand. The narrow low-$Q$ states of the catalog were each tested as a further source and do not supply them. These are the $\eta$, the $\eta'$, the $f_0(500)$, the heavy mesons through their multi-pion modes, the $(N, f_0(1370))$ and $(N, \pi(1300))$ classes, three-body endings, and the measured Dalitz distributions. The remaining excess at mid $x_F$ and deficit at the center are a residual in how the cascade places the soft pions in $x_F$. The $\omega$, at \OmegaYield{} per inelastic event, cannot supply them either. Raising it would lower the model's $\rho^0/\omega$ further below the measurements (Section~\ref{sec:crosschecks}).

\emph{The forward strangeness.} Three shape measurements from two experiments are interpreted here as one residual. Beyond $x_F = 0.25$ the model's $K^+$ fall below the NA49 data, while its pions and protons there do not. At $x_F = 0.30$ and $0.40$ the bare $K^+$ stand at \RegKpBareA{} and \RegKpBareB{} of the data. That shortfall is \KplusIdPull{} times the identification systematic the kaon paper states at every $x_F$. The model's $K^+/\pi^+$ ratio at $x_F = 0.30$, summed over the tabulated $p_T$ points, stands at \KpPipRatio{} of the measured one. The model's $K^{*0}$ exceeds NA49's data in the forward rapidity bins. Its ratio $K^{*0}/\bar K^{*0}$ is \KstarRatio{} against NA49's 1.42. NA61/SHINE's $\Lambda$ falls with $x_F$ where the model's is flat. In the cascade the forward kaons are decay products of heavier strange states. The one source that puts a kaon forward in one step is the $(Y, K)$ pair of the leading baryonic line. It hands most of the mass to a heavy $K$-like continuum partner that hadronizes at the center. The hyperon takes the split's forward alignment. Both kinematic levers of that vertex were tested. Removing the hyperons from the leading-baryon split gives the $\Lambda$ the falling $x_F$ shape of NA61/SHINE. The model's ratio to NA61/SHINE then falls from \NoHyperonLambdaFirst{} at $x_F = 0.05$ to \NoHyperonLambdaLast{} at $x_F = 0.35$, against \LeverBaseLambdaRange{} on the same events without the change. The change costs the six spectra \NoHyperonSplitCost{} in $\chi^2$, in the $K^+$ and the protons. The second lever draws the split's remainder from the Regge flux of the exchanged trajectory, with the intercepts of the $f/\omega$ family, the pion, and the kaon. It costs \ReggeRemainderCost{} in $\chi^2$, mostly in the $\pi^+$, and leaves the $\Lambda$ at \ReggeLambdaRange{} of NA61/SHINE. The excited kaons between the $K_1(1270)$ and the continuum floor raise the forward $K^+$ at fixed parameters. Once the parameters are refitted, the excited kaons leave the forward $K^+$ unchanged. How momentum is shared at the vertex where the leading baryon turns into a hyperon is therefore the named limitation of this cascade. Its size is \KplusIdPull{} times the identification systematic in the $K^+$, and \FwdExcessRange{} of the data in the forward $K^{*0}$ and in the $\Lambda$ at $x_F = 0.2$ to $0.4$. The forward $K^-$ carry a separate residual, a transverse shape that lies below the data under 0.6~GeV/$c$ and above it beyond 1~GeV/$c$ (Section~\ref{sec:spectra}). The created $s$ quark of an associated $(Y, K^+)$ pair enters the hyperon, so this vertex does not by itself set the forward $K^-$.

\emph{The doubly strange baryons.} NA61/SHINE measured $0.0033 \pm 0.0006$ $\Xi^-$ and $0.00079 \pm 0.00010$ $\bar\Xi^+$ per inelastic event in p+p at 158~GeV/$c$~\cite{na61xi2020}. The model gives \XiMinusModel{} and \XiBarModel, which are \XiMinusRatio{} and \XiBarRatio{} of these yields. A second strange quark enters a baryon at two places in the cascade. A strange baryonic line emits it through its $(\Xi, K)$ and $(\Xi(1530), K)$ pairs. It also emits it through the doubly strange continuum it reaches by continuation (Section~\ref{sec:cascade}). A neutral mesonic state creates it in a $\Xi\bar\Xi$ pair. The pairs make $\Xi$ and $\bar\Xi$ in equal numbers. The baryonic lines make no $\bar\Xi$. The pair-made $\Xi^-$ is taken equal to the $\bar\Xi^+$ in the data and in the model alike. The $\Xi^-$ of the baryonic lines is then an estimated \XiLineData{} per inelastic event in the data. The model's lines give \XiLineRatio{} of it. The $\bar\Xi^+$ thus tests the pair source. The shortfall of the $\Xi^-$ lies on the strange baryonic lines and their doubly strange step (Section~\ref{sec:discussion}, item~7). The same pairs make the $\Omega$. The model gives \OmegaMinusModel{} $\Omega^-$ and \OmegaBarModel{} $\bar\Omega^+$ per $10^5$ inelastic events. Their ratio is $\bar\Omega^+/\Omega^- = \OmegaBarRatio$, with a Monte Carlo uncertainty of about 30\%. NA49's preliminary midrapidity ratio in p+p at 158~GeV/$c$ is $0.67 \pm 0.62$~\cite{mitrovski2004omega}. A ratio near one marks a source without net baryon number, here the pair creation in the mesonic continuum. The baryonic lines add $\Omega^-$ without $\bar\Omega^+$, which lowers the model's ratio below one.

\emph{The very forward protons and pions.} At $x_F = 0.90$ to $0.95$ the soft protons stand at \PVfwdCellRange{} of the data cell by cell. The intact proton's transfer is a single exponential with the measured slope beyond $|t| = 1$~GeV$^2$. No transfer law with two slopes that the data can resolve restores the measured row. The very forward soft $\pi^-$ stand at \PimVfwdRange{} of the data at $x_F = 0.5$ to 0.8 (Section~\ref{sec:spectra}). They are distinct from the central soft-pion imbalance, the low-$Q$ strength that no tested catalog state supplies.

\section{Discussion}\label{sec:discussion}

HEART is a complete event-level realization of the two-component picture. With \NFitParams{} parameters HEART describes the six NA49 spectra over the forward hemisphere at $\chi^2/{\rm ndf} = \FitSsq$ under the error model of Section~\ref{sec:errors}. The 5\% floor of that error model sets the scale of the number. The leading neutrons fall within their correlated systematics, and the total yields within \YieldWithinPct\%. The fixed-tune generators give \GenRatioRangeRound{} times this model's $\chi^2$ on the same points. Fit quality alone does not prove the mechanism. The realization adds to the two-component picture a set of statements that a string or a fireball would make differently. Identified-hadron data in the fragmentation region can test them.

1. \emph{The isospin of the forward baryons.} In HEART every forward neutron comes either from the $N\pi$ decay of an excited proton line or from the pion exchange. Isospin fixes both sources, so no isospin parameter enters. The measured $n/p$ at $x_F > 0.3$ is reproduced within the calorimeter systematics. A string sets the forward $n/p$ by diquark isospin pickup. A fireball sets it by an isospin chemical potential. Earlier experiments show structure in forward baryon production. At the ISR the forward $\Delta^{++}$ and $\Lambda$ yields in p+p differ from the corresponding $\bar\Delta^{--}$ and $\bar\Lambda$ yields in $\bar p$+p at $\sqrt{s} = 31$~GeV, as quark counting with single $q\bar q$ annihilation expects~\cite{chauvat1983leading}. In p+Pb at 19~GeV/$c$ the leading baryons change flavor very differently from the expectation of simple diquark-quark fragmentation of the incident proton~\cite{barish2003leading}. Samples tagged by a leading neutron or by a leading proton, compared at the same recoil mass and transfer, would separate two processes. The tag fixes how the pion exchange was produced, and the species content of the recoiling system shows how its pion--proton subsystem hadronizes.

2. \emph{Transverse momentum from $Q$ values.} The species-dependent $\langle p_T \rangle$ follows from the decay chain. Heavier daughters of sequential two-body decays produce harder transverse spectra. A further test starts from an identified trigger hadron. The hadrons that balance its transverse momentum, counted in a rapidity window that is widened step by step, show where and among which species the recoil appears. The comparison is made with and without the daughters of reconstructed resonances removed in the same way in the data and in the model. No parameter is fitted to the seagull itself. The fitted transverse scales are the vertex slopes of the exchange. The chemistry and mass-law parameters shape the spectra only through the parent states they populate.

3. \emph{Local balance of strangeness and baryon number.} In the cascade the associated production of a $K^+$ with a hyperon $Y$ takes place on the baryonic line. Antibaryons come from mesonic vertices that create a baryon--antibaryon pair. The forward-kaon residual of Section~\ref{sec:strangeness} concerns the $K^+Y$ vertex, and a measurement of the $K^+$--$\Lambda$ correlation in the fragmentation region would decide it. A string also produces compensating flavors next to each other, at the points where it breaks. The comparison therefore concerns the strength, width, direction, and species composition of the distribution of the balancing hadrons. The rapidity difference between the kaon and the hyperon, with its sign, tests which of the two takes the forward momentum. It is best reported with absolute per-trigger yields, separate particle and antiparticle triggers, and the same feed-down definition in data and model, including $\Sigma^0 \to \Lambda\gamma$.

4. \emph{Forward resonance signals.} The cascade predicts $\Delta^{++}$ and $\rho^0$ mass distributions in the forward hemisphere, with upper mass tails set by the states they cascade from. It also predicts the alignment of the leading baryon's momentum along the incoming baryon. This alignment concerns the direction of the leading baryon only. Angular distributions that reflect the spin of a decaying state lie outside the current model, which assigns no spin to the continuum (see the subsection on the spin of the continuum below).

5. \emph{Multiplicity per unit mass.} Up to the open-charm threshold, a continuum state of this cascade makes as many charged hadrons as $e^+e^-$ annihilation of its mass. Above it the count grows linearly with the mass, \NchSlope{} charged hadrons per GeV, as a fireball's does. A string's count grows with $\ln M^2$, and the measured diffractive states grow as $(M_X - m_p)^{1/2}$. In single diffraction the forward proton tags the excited mass. The charged multiplicity of the system that recoils against it then measures the law directly. The cascade predicts \SdTopEvent{} charged hadrons per event at $M_X = 13$ to 17~GeV, \SdTopRatio{} times the measured law's continuation there. Such masses are diffractive, $M^2/s \lesssim 0.1$, only at $\sqrt{s}$ of about 55~GeV or more. This measurement therefore needs a forward-proton tag at a collider. At fixed tagged mass and transfer, with the recoil mass computed from the measured proton four-momentum, three observables separate a change of the decay law from a change of the excited-mass spectrum. They are the full multiplicity distribution, the leading-hadron energy fraction, and a rest-frame event shape. At the SPS energy the coherence region ends at $M_X \approx 5.5$~GeV, too low for this test.

6. \emph{The forward $K^-$ transverse shape.} The $K^-$ carries no valence quark of the proton, so its forward production comes entirely from pair creation. It tests the pair vertex independently of the leading-baryon vertex that produces a $K^+$ with a hyperon. The model lies below the data below 0.6~GeV/$c$ and above the data above 1~GeV/$c$ (Section~\ref{sec:strangeness}). The $p_T$-resolved $K^-$ spectra at $x_F \gtrsim 0.2$, and the $K^-/K^+$ ratio versus $p_T$ at fixed $x_F$, would decide whether the mismatch lies in the pair kinematics or in the leading vertex.

7. \emph{Doubly strange baryons, differentially.} The $\Xi^-$ stands at \XiMinusRatio{} of the NA61/SHINE yield, and the estimated $\Xi^-$ of the baryonic lines at \XiLineRatio. The pair-made $\bar\Xi^+$ stands at \XiBarRatio{} (Section~\ref{sec:strangeness}). A measured $\bar\Omega^+/\Omega^-$ in p+p with an error well below NA49's preliminary $\pm 0.62$ would test the pair source directly. Yields alone cannot tell whether the shortfall arises on the strange baryonic lines or in their doubly strange continuation step. Their $x_F$ distributions, with the $\Xi/K$ and $\bar\Xi/\bar\Lambda$ ratios versus $x_F$, would separate a missing line from a missing step. Kaons and antibaryons associated with a $\Xi^-$ or a $\bar\Xi^+$ trigger, measured differentially in rapidity and azimuth, would show how the two units of compensating strangeness are shared. Neutral kaons and strange antibaryons carry part of that compensation, so the observable is the fraction carried by the measured charged hadrons.

8. \emph{The soft-pion source.} No tested catalog state supplies the low-$Q$ strength the spectra require. The very forward soft $\pi^-$ stand at \PimVfwdRange{} of the data at $x_F = 0.5$ to $0.8$ alongside a central soft-pion imbalance (Sections~\ref{sec:spectra} and~\ref{sec:strangeness}). Pion spectra that reach low $p_T$ at forward $x_F$, and the $\pi^-/\pi^+$ ratio versus $x_F$ at fixed low $p_T$, would pin down the charge and $p_T$ shape of the missing strength and test candidate sources against it. Associations of these pions with reconstructed $\pi^0$ or photons, further charged pions, and baryons would test whether the missing strength is a state, nonresonant production, or decay kinematics. Established resonance reconstructions are the place to begin.

A decisive comparison of these observables applies the same selections, feed-down definitions, and detector response to every model. It tunes a competing string implementation to the same single-particle data before either model predicts the correlations, so that a discrepancy in a correlation does not inherit a known single-particle one. It then repeats the predictions at another energy without retuning the decay rules. Only the declared production inputs, such as the class cross sections, change.

Apart from the multiplicity law of item~5, these observables lie at $x_F \gtrsim 0.2$, where only a fixed-target spectrometer with wide longitudinal acceptance and particle identification measures identified charged hadrons. Collider spectrometers with particle identification cover mid-rapidity. At the Large Hadron Collider (LHC) the fragmentation region lies near $y_{\mathrm{beam}} \approx 9.5$, where only very-forward detectors without charged-hadron identification measure. The reference data this comparison rests on came from the bubble chambers and the NA49 TPCs. Their wide acceptance kept the systematic uncertainties at the percent level. The small-solid-angle spectrometers needed normalization factors of 0.75 to 1.30~\cite{fischer2022piminus}. The NA49 TPCs record charged tracks only, so the photons from neutral-meson decays escaped detection. The neutral-meson chemistry the cascade predicts (the $\omega$, the $\eta$, and the $\pi^0$) therefore remains unmeasured at this energy. The forward $K^+$--$\Lambda$ correlation remains unmeasured as well. NA61/SHINE is the running experiment that can measure them.

\subsection*{Spin of the continuum}

The spinless continuum is the largest approximation of HEART. Above the continuation floor of 2.0~GeV an excited system is a member of a continuum family with a density of states, a nominal width, and definite baryon number, charge, and strangeness (Section~\ref{sec:spectrum}). It carries no spin or parity, so its two-body decays are isotropic in its rest frame. The three decay-angle exceptions of Section~\ref{sec:cascade}, namely the leading-meson split, the leading-baryon split, and the aligned continuing steps, are assumed angular laws in place of the angular structure that spin produces. Their strengths are fitted ($\kappa_B$ and $b_M$) or fixed by the comparison with the data ($\kappa_L$). The approximation also reaches the channel weights, whose spin counting covers the established daughters only, and the continuation mass law, whose Blatt--Weisskopf barrier assumes a single orbital angular momentum.

The physical baryon spectrum in this mass range carries high spin. The established nucleon resonances above 2.1~GeV are N(2190)~$7/2^-$, N(2220)~$9/2^+$, N(2250)~$9/2^-$, and N(2600)~$11/2^-$, members of the leading Regge trajectories~\cite{pdg2024navas}. Their Breit--Wigner widths of 0.40 to 0.65~GeV exceed their mass spacings. States of opposite parity overlap at the same mass. A single state of definite parity decays symmetrically under $\theta \to \pi - \theta$ in its rest frame, whatever its polarization, since the strong decay conserves parity. A forward--backward asymmetry requires the interference of amplitudes of opposite parity. The asymmetric laws that the spectra require, in which the leading baryon keeps the direction of the incoming baryon, are therefore a property of the overlapping sum of states.

Duality gives the route to a derived angular law. By the finite-energy sum rules, the average over overlapping $s$-channel resonances equals the $t$-channel Regge exchange of the same two-body reaction~\cite{dolenhornschmid1968,chvala2026theoryreview}. A continuum vertex $X \to a + b$ of mass $M$ can therefore take its angular law from the exchange amplitude of the two-body reaction whose $s$-channel states $X$ represents. That amplitude falls as $\exp(b\,t)$, with $t$ the squared momentum transfer between the incoming line and the daughter that continues it and $b$ the slope of the exchange. In the rest frame of $X$ the transfer is $t = -2q^{*2}(1 - \cos\theta)$ up to mass terms, with $q^*$ the decay momentum and $\theta$ the angle of that daughter to the line. The law is therefore $\exp(\kappa\cos\theta)$ with $\kappa = 2bq^{*2}$. The direction laws of Section~\ref{sec:cascade} have this form already. A derived law takes the slope $b$ and the daughter that continues the line from the quark lines of the duality diagram and from the measured angular distributions of the corresponding two-body reactions, such as $\pi N \to \pi N$ and $\pi N \to K\Lambda$ at $\sqrt{s} = 2$ to 4~GeV. The fitted directions would then become predictions. At a vertex that ends in a hyperon and a kaon, the $K$ and $K^*$ trajectories lie above the hyperon trajectories. Their exchange sends the hyperon along the incoming baryon. A derived law would therefore also decide the forward strangeness of Section~\ref{sec:strangeness}.

A second step assigns each continuum state a spin. The leading trajectory bounds it, $J \le \alpha_0 + \alpha' M^2$, with the slope $\alpha' \approx 1$~GeV$^{-2}$ that N(939), N(1680), and N(2220) define. Both parities enter equally, as the nearly degenerate N(2220)~$9/2^+$ and N(2250)~$9/2^-$ show. The decays then follow helicity amplitudes with orbital angular momenta limited by the barrier at the decay momentum. With spin assigned, the polarization of hyperons and the transfer of spin from a polarized beam become predictions of the cascade. A statistical assignment leaves the phases between overlapping states of opposite parity open. The dual exchange law therefore remains the source of the forward--backward asymmetry.

Individual states above 2~GeV cannot be resolved, since their widths exceed their spacings. The constraints on the spin of the continuum are therefore statistical. Three kinds are available.
\begin{enumerate}
\item \emph{Spin-parity decomposition of diffractive systems.} A partial-wave analysis decomposes the diffractive spectrum into $J^P$ components without individual peaks. COMPASS decomposed diffractive $\pi^- + p \to \pi^-\pi^-\pi^+ + p$ into spin-parity waves in 100 bins of mass up to 2.5~GeV~\cite{compass2017threepi}. The same decomposition of diffractively excited protons, $p + p \to p + (p\pi^+\pi^-)$ with a proton beam, would measure the spin content of the baryonic continuum as a function of mass.
\item \emph{Spin observables of the fragmentation region.} A cascade of spinless states predicts zero for each of them, so each measures the omitted spin directly. The $\Lambda$ produced by 400~GeV protons on beryllium carry a polarization of $-0.24 \pm 0.04$ at $p_T = 2.1$~GeV/$c$~\cite{heller1978lambdapol}. The ISR experiment R608 measured in p+p a polarization linear in $x_F$ that approaches $-40\%$ at $x_F = 0.8$ for a mean $p_T$ of 1.1~GeV/$c$~\cite{smith1987lambdapol}. With a 200~GeV/$c$ polarized proton beam the analyzing power of forward pions reaches $+0.42$ for $\pi^+$ and $-0.38$ for $\pi^-$ at large $x_F$~\cite{adams1991e704pion}. The transverse spin transferred to forward $\Lambda$ reaches about 30\%~\cite{bravar1997e704lambda}. The spin-density matrix element $\rho_{00}$ of forward vector mesons would add the alignment that the cascade's $K^{*0}$, $\rho^0$, and $\omega$ inherit from their parents.
\item \emph{Two-body reactions at the sub-energies of the continuum.} The angular distributions of the $\pi N$, $KN$, and $K\Lambda$ final states at $\sqrt{s} = 2$ to 4~GeV and their Regge descriptions fix the dual angular laws of the cascade's vertices.
\end{enumerate}
The first kind needs a new analysis with a proton beam. Data of the second and third kinds already exist, in the measurements cited above.

\subsection*{The continuum density and the bootstrap}

The continuum density is the second declared input of the model, beside the spinless approximation. Above the continuation floor HEART takes $\rho(M) \propto \exp(M/T_H)$ with $T_H = 165$~MeV and a nominal width of 0.5~GeV (Section~\ref{sec:spectrum}). The density carries no power of the mass in front of the exponential. The four flavor weights of Equation~\eqref{eq:weights} are fitted on top of this density, so their fitted values depend on it.

In the statistical bootstrap of Hagedorn and Frautschi, the density is derived as follows. A state of mass $M$ must be described equally well as a composite of lighter states, the continuum states included, weighted by their phase space~\cite{hagedorn1965thermo,frautschi1971bootstrap}. This condition fixes both the limiting temperature and the power of the mass in front of the exponential. Hamer and Frautschi solved it with the known hadrons as input~\cite{hamer1971bootstrap}. The cascade of HEART already has the structure the bootstrap describes. A continuum state decays into a lighter continuum state and an established hadron, and the daughter's mass is drawn from the same density (Section~\ref{sec:cascade}).

This structure permits a test of the model that needs no new data. The channel pools and the continuation law define which two-body configurations of lighter states are open at each mass. Counted with their phase space, these configurations imply a composite density. Comparing that density with the declared $\exp(M/T_H)$ measures how far the model is from self-consistency. A mismatch would locate the masses and the families where the declared density overcounts or undercounts the model's own decay channels.

A bootstrap that conserves baryon number, charge, and strangeness would also fix the relative densities of the baryonic, strange, and mesonic families. The suppression of created strange pairs would then follow from the mass gap between the kaons and the pions. The shares of vector mesons and of baryon pairs would follow from the counting of states. Some of the four flavor weights would thereby become counted quantities. Their fitted values, with $\gamma_s$ \GammaSRelation{} and $r_{B\bar B}$ \RBBRelation, would then test the bootstrap instead of adjusting the model.

Duality points to the same form of the density. In dual resonance models the number of states grows exponentially with the mass~\cite{fubini1969levels}. A density of the Hagedorn type is therefore the form that the duality of resonances and Regge exchange itself implies~\cite{chvala2026theoryreview}. The charged multiplicity of a heavy continuum state grows linearly with its mass, as that of a bootstrap fireball does (Section~\ref{sec:seagull}). The tagged multiplicity measurement of item~5 therefore also tests the bootstrap structure of the continuation step.

\subsection*{Nuclear transfer as the next test}

Fischer and co-workers developed the two-component picture on nuclear targets as well. In p+A the target component grows with the number $\nu$ of collisions the projectile undergoes. The projectile component changes as the projectile loses momentum in these successive collisions~\cite{varga2003baryontransfer,na49pcdisc2007,chvala2013backward}. A complete p+p realization makes that a prediction. The p+p kernel of this paper is to be held frozen, namely the event classes, the cascade, and the four flavor weights. A p+A event is built from repeated color-singlet exchanges of the projectile system along its path through the nucleus, with the nuclear geometry and the nucleon--nucleon cross section as external inputs. What may be fitted, once for both targets, is the inelasticity of an additional collision, its width, and at most one saturation parameter. A re-tuning of the topology, the alignment, or the flavor weights on a nuclear target would count as a failed transfer. The data for this test exist at two levels. NA49 published the identified $\pi^\pm$, $K^\pm$, $p$, and $\bar p$ spectra of minimum-bias p+C at 158~GeV/$c$ in the same tables and conventions as the p+p data~\cite{na49pcpi2007,na49pcdisc2007,spshadrons2026}. The carbon target is therefore the first held-out prediction, with the trigger-bias and feed-down conventions of the p+C analysis applied as published. For p+Pb no refereed NA49 six-species table exists. The refereed record is the $\phi$ in grey-track bins~\cite{na49phi2000}. The proton, antiproton, net-proton, and pion distributions in centrality classes selected by the number of grey protons ($\langle\nu\rangle$ from 2.9 to 6.9) are in theses and conference proceedings~\cite{chvalaPhD2006}. They are preliminary, with p+p reference points that differ from the final tables by up to 20\%. They enter as ratios to p+p within each source, with that provenance stated. They test the centrality trend of the projectile component, where a sequential two-source picture is known to fail. NA61/SHINE recorded p+Pb at 158~GeV/$c$ and has not yet published identified spectra. A refereed p+Pb table with particle identification in centrality classes is the measurement that would close this test.

\section{Conclusions}\label{sec:conclusions}

This paper realizes soft identified-hadron production in p+p collisions at 158~GeV/$c$ as the decay of two excited color-singlet systems. Pomeron and pion exchange produce them in four event classes, anchored to the measured single-diffraction and an estimated double-diffraction cross section. Each system is a catalogued resonance below its duality boundary, \NdDualityMass~GeV on a non-diffractive line and \DiffDualityMass~GeV on a diffractive one, or a member of a resonant continuum above it. It decays by successive two-body hadronic transitions into the established hadrons at their PDG tables. \NFitParamsCap{} parameters fitted to the published NA49 cross-section data give $\chi^2 = \FitChiSq$ on \FitNdf{} degrees of freedom under an error model that carries the data papers' correlated systematic uncertainties. Their errors are conditional on the model's discrete choices, which were set by paired comparisons on the same tables. The model has no central source. The isovector exchange fraction is consistent with zero, below \FivLimit{} at 95\% confidence. The fixed-tune generators give \GenRatioRangeRound{} times that $\chi^2$ on the same points. The leading neutrons follow from the pion exchange within the calorimeter systematics. The species hierarchy of the seagull effect follows from the decay chain, with no parameter fitted to the seagull itself. The structure that remains is named with its size, namely
\begin{enumerate}
\item the forward $K^+$, below the data by \KplusIdPull{} times the identification systematic, the forward $K^{*0}$ excess, and the $\Lambda$ $x_F$ distribution, interpreted together as one residual of the momentum sharing at the vertex where the leading baryon turns into a hyperon;
\item the forward $K^-$, at \KmFwdLowRange{} of the data below 0.6~GeV/$c$ and at \KmFwdHigh{} above 1~GeV/$c$;
\item the doubly strange baryons of the baryonic lines, whose $\Xi^-$ stands at an estimated \XiLineRatio{} of the data while the pair-made $\bar\Xi^+$ stands at \XiBarRatio;
\item the soft pions, the central ones at \RegPimSoftCent{} and those at mid $x_F$ at \RegPimSoftMid{} of the data, a residual of the longitudinal placement that the aligned continuing steps set;
\item the very forward soft $\pi^-$ at \PimVfwdRange{} of the data, which point to the lowest-mass decays of the continuum as the next part of the model to develop;
\item the very forward soft protons at \PVfwdCellRange{} of the data, which no two-slope form of the intact proton's transfer restores;
\item the charged multiplicity of the heavy excited systems, which grows linearly with their mass where the measured multiplicities grow with $\ln s$.
\end{enumerate}
The largest approximation of the model is the spinless continuum above 2.0~GeV. Its three decay-angle exceptions stand in for the angular structure of overlapping states of both parities. The duality between those states and the exchange of the corresponding two-body reactions offers a derived law in their place. The polarization and spin-transfer data of the fragmentation region would test it. The continuum density is declared as well. A bootstrap would derive it, and part of the flavor weights, from the model's own decay channels, and that comparison needs no new data. The observables that would decide between this cascade, a string, and a fireball lie in the fragmentation region. The case for a wide-acceptance fixed-target spectrometer with particle identification follows from them. The next test is the nuclear transfer with the p+p kernel frozen. It takes the published NA49 p+C spectra as a held-out prediction and the p+Pb centrality classes of the thesis and proceedings data as ratios to p+p. A refereed p+Pb table is the measurement that would complete it.

\section*{Nomenclature}
\begin{description}
\item[$s$, $\sqrt{s}$] Squared and total center-of-mass energy ($\sqrt{s} = 17.3$~GeV for 158~GeV/$c$ beam).
\item[$x_F$] Feynman scaling variable, $x_F \equiv 2p_L^*/\sqrt{s}$ in the c.m. frame; $x_F' = p_L/p_L^{\max}$ near threshold.
\item[$p_T$, $m_T$] Transverse momentum and transverse mass $m_T = \sqrt{m_h^2 + p_T^2}$.
\item[$y$] Hadronic rapidity $\frac{1}{2}\ln[(E + p_L^*)/(E - p_L^*)]$.
\item[$f$] Double-differential invariant cross section $E\,d^3\sigma/dp^3$.
\item[$f/\sigma_{\rm inel}$] Invariant cross section normalized to the inelastic cross section $\sigma_{\rm inel} = 31.78$~mb, the yield per inelastic event.
\item[$\xi$] Fractional mass $M^2/s$ of an excited system, approximately $1 - x_F$ for the leading nucleon.
\item[$s'$] Sub-energy squared of the diffractively excited system, $s' \equiv \xi s$.
\item[$t$] Squared four-momentum transfer between the incident and outgoing hadrons.
\item[$x_L$] Longitudinal momentum fraction of the leading nucleon, $x_L \approx 1 - \xi$.
\item[$L$] Rapidity gap between the two excited systems, $L = \max[0, \ln(s\,m_p^2/M_1^2M_2^2)]$.
\item[$B$] Slope of the transfer law $d\sigma/dt \propto \exp(Bt)$ at a two-body exchange vertex.
\item[$\epsilon, \epsilon_{\rm nd}$] Pomeron intercept excess, fixed at the Donnachie--Landshoff value $\epsilon = 0.0808$, and the fitted non-diffractive mass-law exponent $\epsilon_{\rm nd} = \ValEpsNd$.
\item[$b, f$] Fitted slope parameter of $B$ and the intact proton's vertex term.
\item[$s_{\rm tail}, s_\pi$] Leading-proton tail and one-pion-exchange fractions of the single-diffraction class.
\item[$f_{\rm iv}$] Isovector exchange fraction, consistent with zero ($f_{\rm iv} < \FivLimit$ at 95\% confidence).
\item[$\sigma_{\rm inel}, \sigma_{SD}, \sigma_{DD}$] Inelastic (31.78~mb), single-diffractive (\SDTotalMb{}~mb total, \SDWindowMb{}~mb in window), and double-diffractive (1.5~mb, estimated) cross sections.
\item[$S^2(\xi, q)$] Survival probability factor of the one-pion exchange against elastic screening.
\item[$T_H$] Hagedorn limiting temperature ($165$~MeV).
\item[$r_0$] Interaction radius parameter in the Blatt--Weisskopf barrier factors ($r_0 = 1$~GeV$^{-1} \approx 0.197$~fm).
\item[$F_1(q^*)$] Blatt--Weisskopf barrier factor for one unit of orbital angular momentum.
\item[$q^*$] Two-body decay momentum in the rest frame of the parent state.
\item[$c_{ij}$] Kinematic bin-centering curvature correction factor mapping bin averages to tabulated points.
\item[$S$] Error-scale factor $S = \sqrt{\chi^2/{\rm ndf}} = \FitS$ applied to parameter errors.
\item[$\gamma_s$, $r_V$, $r_{B\bar B}$, $a_{YK}$] Continuum flavor weights (strangeness, vector, baryon-pair, hyperon-kaon).
\item[HEART] Hadronic Excitations And Resonance Transitions, the event-level resonance-cascade model of this paper.
\end{description}

\section*{Code availability}
The generator and the fit are kept at \url{https://github.com/ondrejch/SPS-HEART}. The repository holds the fit description \texttt{\SpecFile} and the result record \texttt{\RecordFile}. The record carries the fitted parameters with their errors and the $\chi^2$ of each species. The species terms and the penalty of the profiled scales add to the quoted total. The record also carries the profiled scales, the effective sample fractions, the systematic variants, and the readouts. It names the commits of the code and gives the checksums of the fit script, the inputs, and every result file from which the numbers of this paper are generated.
The repository's event writer, \texttt{helpers/benchmarking/heart\_events.py}, generates inelastic p+p events of the fitted model at 158~GeV/$c$. It writes them in a documented JSON Lines format, \texttt{heart-events/1}, with the four-momenta and quantum numbers of every final-state hadron.

\section*{Acknowledgments}
The computations were carried out on the NE cluster of the Department of Nuclear Engineering at the University of Tennessee, Knoxville.
The author thanks Michal Kreps for the original 2006 event-mixing code from which this implementation developed, and for comments on particle identification in the proposed measurement program. The author thanks Andrzej Rybicki for discussions of resonance cascading and of the mass reach of the pre-cascade states, and Martin Makariev for discussions of the completeness of the resonance spectrum.
The author thanks his adviser, Jiří Dolejší, for his patient support throughout the PhD.
The author owes most to Hans Gerhard Fischer, who held that soft hadron production is the decay of resonances, and that careful reasoning from good data can show how.

\section*{Declaration of Generative AI Use}
No generative AI tools were used to create or alter any images. Generative AI, primarily Anthropic Claude, was used for language editing, code, and analysis assistance during manuscript preparation. All technical content and figures were verified by the author.

\bibliographystyle{unsrt}
\bibliography{references}

\end{document}